\documentclass[amsmath,amssymb,a4paper,prc,final,superscriptaddress,
showpacs,twocolumn]{revtex4} 

\usepackage[T1]{fontenc}
\usepackage[latin1]{inputenc}
\usepackage{bm,hyphenat,xspace}
\usepackage{epsfig,pstricks,psfig,epic,eepic,pst-coil}

\newcommand{\scl}{0.6}

\newcommand{\Eq}{Eq.}
\newcommand{\Eqs}{Eqs.}
\newcommand{\Fig}{Fig.}
\newcommand{\Figs}{Figs.}
\newcommand{\Ref}{Ref.}
\newcommand{\Refs}{Refs.}
\newcommand{\Sect}{Sec.}
\renewcommand {\vec}[1]{{\mathbf{#1}}}
\newcommand {\vecg}[1]{\mbox{\boldmath{$#1$}} }
\newcommand {\Kpl}{\vec{K}_{+}}

\newcommand {\Ei}{E_i \! + \!i0}
\newcommand{\mhk}{\hat{\mathbf{k}}}
\newcommand{\kgl}{\mathbf{k}_{\gamma}}
\newcommand{\kgc}{\mathbf{k}_{\gamma \, \mathrm{cm}}}

\begin{document}

\title {Trinucleon photonuclear reactions with $\Delta$-isobar excitation: \\
Processes below pion-production threshold}

\author{A.~Deltuva} 
\thanks{On leave from Institute of Theoretical Physics and Astronomy,
Vilnius University, Vilnius 2600, Lithuania}
\email{deltuva@itp.uni-hannover.de}
\affiliation{Institut f\"ur Theoretische Physik,  Universit\"at Hannover,
  D-30167 Hannover, Germany}

\author{L.~P.~Yuan} 
\affiliation{Institut f\"ur Theoretische Physik,  Universit\"at Hannover,
  D-30167 Hannover, Germany}

\author{J.~Adam~Jr.} 
\affiliation{Nuclear Physics Institute, CZ-25068 \v{R}e\v{z} near Prague, 
Czech Republic}

\author{A.~C.~Fonseca} 
\affiliation{Centro de F\'{\i}sica Nuclear da Universidade de Lisboa, 
P-1649-003 Lisboa, Portugal }

\author{P.~U.~Sauer}
\affiliation{Institut f\"ur Theoretische Physik,  Universit\"at Hannover,
  D-30167 Hannover, Germany}
\received{13 August 2003}


\begin{abstract}
 Radiative nucleon-deuteron capture and  two- and three-body photo 
disintegration of the three-nucleon bound state are described. The description
 uses the purely nucleonic charge-dependent CD-Bonn potential and
its coupled-channel extension CD Bonn + $\Delta$. The $\Delta$-isobar
excitation yields an effective three-nucleon force and effective two- and
three-nucleon currents besides other $\Delta$-isobar effects; they
are mutually consistent. Exact solutions of three-particle  
equations are employed for the initial and final states of the reactions. 
The current has one-baryon and two-baryon contributions and couples
nucleonic with $\Delta$-isobar channels.
$\Delta$-isobar effects on the observables are isolated. Shortcomings
of the theoretical description are  discussed and their consequence for the
calculation of observables is estimated.

\end{abstract}

 \maketitle

\section{Introduction} \label{sec:intro}

Photo reactions  in the three-nucleon system are described. The available
scattering energy stays below  pion-production threshold. The description
allows for the excitation of a nucleon to a $\Delta$ isobar. 
The excitation of the $\Delta$ isobar remains virtual due to the available
energy. The $\Delta$ isobar is therefore considered a stable particle;
it yields an effective three-nucleon force and effective exchange currents
besides other $\Delta$-isobar effects. The effective three-nucleon force 
simulates the two-pion exchange Fujita-Miyazawa force~\cite{fujita:57a} and the 
three-pion ring part of the Illinois forces~\cite{pieper:01a} in a reducible
energy-dependent form. The effective exchange currents are of two-nucleon
and three-nucleon nature. Since the effective nucleonic forces and currents
are built from the same two-baryon coupled-channel potential and from the 
corresponding one-baryon and two-baryon coupled-channel current, they are consistent
with each other. Since the two-baryon coupled-channel potential is based on 
the single exchanges of the standard isovector mesons pi $(\pi)$ and
rho $(\rho)$ and of the isoscalar mesons omega $(\omega)$ and sigma $(\sigma)$,
the same meson exchanges are contained in the effective nucleonic forces 
and in the effective nucleonic currents. 
E.g., besides the $\pi$ exchange of \Refs~\cite{fujita:57a,pieper:01a} 
also $\rho$ exchange is included in forces and currents.

The exact solution of the three-particle scattering equations is used for
the description of the initial- and final-state interactions; the 
coupled-channel formulation  for nucleon-deuteron scattering is
developed in \Refs~\cite{nemoto:98a,nemoto:98b,chmielewski:03a};  
radiative nucleon-deuteron capture and electromagnetic (e.m.) 
two-body breakup of the three-nucleon bound state
are described in \Ref~\cite{yuan:02a}. Whereas a separable expansion
of the Paris potential~\cite{lacombe:80a} is used in those early calculations, 
\Ref~\cite{deltuva:03a} solves the three-particle scattering equations  exactly
by Chebyshev expansion of the two-baryon transition matrix
as interpolation technique; that technique is found highly efficient 
and systematic. In this paper, the technique of  \Ref~\cite{deltuva:03a} is also 
used for the description of photo processes in the three-nucleon system.
In contrast to \Ref~\cite{yuan:02a} the underlying purely nucleonic 
reference potential is CD Bonn~\cite{machleidt:01a}.
Furthermore, the coupled-channel extension of CD Bonn, called CD Bonn + $\Delta$
 and employed in this paper,
is fitted in \Ref~\cite{deltuva:03c} to the experimental two-nucleon
data up to 350~MeV nucleon lab energy; it is as realistic  as CD Bonn.
Thus, this paper updates our previous calculations~\cite{yuan:02a} of 
trinucleon photo reactions. Compared to \Ref~\cite{yuan:02a},
the description is extended to higher energies, and 
 three-nucleon breakup is also included; however,
energetically the description remains below  pion-production threshold.
An alternative description of e.m. processes in the three-nucleon system
is given in \Refs~\cite{golak:00a,skibinski:03a,skibinski:03b};   
 \Refs~\cite{golak:00a,skibinski:03a,skibinski:03b} employ a different
two-nucleon potential, an explicit irreducible three-nucleon force and a
different e.m. current; nevertheless, the theoretical predictions of 
 \Refs~\cite{golak:00a,skibinski:03a,skibinski:03b} and of this paper
will turn out to be qualitatively quite similar.

Section~\ref{sec:calc} recalls our calculational procedure and especially
stresses its improvements. Section~\ref{sec:res} presents characteristic 
results for observables; $\Delta$-isobar effects on those observables are
isolated. Section~\ref{sec:shortcomings} discusses the technical shortcomings
of the given results.
 Section~\ref{sec:concl} gives a summary and our conclusions.

\section{\label{sec:calc} Calculational procedure}

The calculational procedure, including the notation, is taken over from
\Ref~\cite{yuan:02a}. We remind the reader shortly of that procedure in order
to point out  changes and to describe the extension to three-body photo
disintegration, not discussed in \Ref~\cite{yuan:02a}.

\subsection{\label{sec:JV} Nonrelativistic model for the electromagnetic
and hadronic interaction of baryons}

\begin{figure}[t]
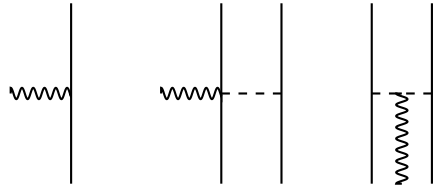

\begin{center}
\pspicture(6.0,3.0)
\def\nucleon{\psline(0,0)(0,2.4)}
\def\NNNN{\psline(0,0)(0,2.4)\psline(0.8,0)(0.8,2.4) 
\psline[linestyle=dashed,dash=3pt 3pt](0,1.2)(0.8,1.2)}
\def\photon{\pscoil[coilwidth=0.15cm,coilaspect=0,coilarm=0.0cm](0,0)(0.8,0)}
\rput(1,0){\nucleon}
\multips(3,0)(2,0){2}{\NNNN}
\multips(0.2,1.2)(2,0){2}{\photon}
\pscoil[coilwidth=0.15cm,coilaspect=0,coilarm=0.0cm](5.4,0)(5.4,1.2)
\endpspicture
\end{center}
\caption{ \label{fig:Jnn} 
 One- and two-baryon processes contained in the used e.m. current. 
In this figure only the purely nucleonic processes are depicted;
the nucleon is indicated by the thin solid line, the photon by the wavy line,
and the  instantaneous meson exchange by the dashed line.
In nonrelativistic order the one-nucleon process contributes to the charge density
and to the spatial current, the two-nucleon processes only to the spatial current.
The  diagonal isovector  $\pi$ and $\rho$ exchanges are taken into account 
in the two-nucleon processes as well as  the nondiagonal  $\rho\pi\gamma$ 
and $\omega\pi\gamma$ contributions. }
\end{figure}

\begin{figure}[!]
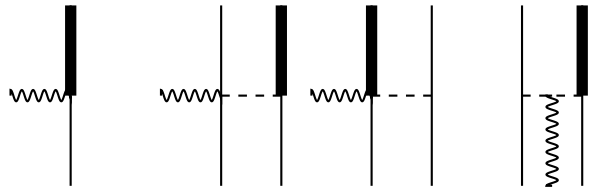

\begin{center}
\pspicture(8.0,3.0)
\def\nucleon{\psline(0,0)(0,2.4)}
\def\NNNN{\psline(0,0)(0,2.4)\psline(0.8,0)(0.8,2.4) 
\psline[linestyle=dashed,dash=3pt 3pt](0,1.2)(0.8,1.2)}
\def\photon{\pscoil[coilwidth=0.15cm,coilaspect=0,coilarm=0.0cm](0,0)(0.8,0)}
\def\deltaiso{\psline[linewidth=0.15cm](0,0)(0,1.2)}
\rput(1,0){\nucleon}
\multips(3,0)(2,0){3}{\NNNN}
\multips(1.0,1.2)(4,0){2}{\deltaiso}
\multips(3.8,1.2)(4,0){2}{\deltaiso}
\multips(0.2,1.2)(2,0){3}{\photon}
\pscoil[coilwidth=0.15cm,coilaspect=0,coilarm=0.0cm](7.4,0)(7.4,1.2)
\endpspicture
\end{center}
\caption{\label{fig:Jnd}
One- and two-baryon processes contained in the used e.m. current. 
In this figure processes are depicted in which one nucleon is turned into a 
$\Delta$ isobar, indicated by a thick line.
 The hermitian adjoint processes are taken into account, 
but are not diagrammatically shown.
In nonrelativistic order the one-baryon and two-baryon processes contribute 
only to the spatial current. 
In the one-baryon current only the magnetic dipole transition is kept.
The  diagonal isovector  $\pi$ and $\rho$ exchanges are taken into account 
in the two-baryon processes as well as  the nondiagonal  $\rho\pi\gamma$ 
and $\omega\pi\gamma$ contributions. }
\end{figure}

\begin{figure*}[!]
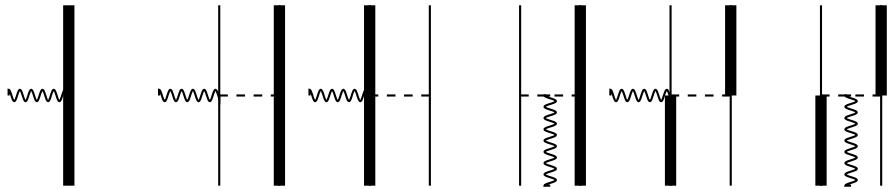

\begin{center}
\pspicture(12.0,3.0)
\def\nucleon{\psline(0,0)(0,2.4)}
\def\NNNN{\psline(0,0)(0,2.4)\psline(0.8,0)(0.8,2.4) 
\psline[linestyle=dashed,dash=3pt 3pt](0,1.2)(0.8,1.2)}
\def\photon{\pscoil[coilwidth=0.15cm,coilaspect=0,coilarm=0.0cm](0,0)(0.8,0)}
\def\deltaiso{\psline[linewidth=0.15cm](0,0)(0,1.2)}
\def\Deltaiso{\psline[linewidth=0.15cm](0,0)(0,2.4)}
\multips(3,0)(2,0){5}{\NNNN}
\multips(1.0,0.0)(4,0){2}{\Deltaiso}
\multips(3.8,0.0)(4,0){2}{\Deltaiso}
\multips(0.2,1.2)(2,0){3}{\photon}
\pscoil[coilwidth=0.15cm,coilaspect=0,coilarm=0.0cm](7.4,0)(7.4,1.2)
\multips(9.0,0.0)(2,0){2}{\deltaiso}
\multips(9.8,1.2)(2,0){2}{\deltaiso}
\multips(8.2,1.2)(2,0){1}{\photon}
\pscoil[coilwidth=0.15cm,coilaspect=0,coilarm=0.0cm](11.4,0)(11.4,1.2)
\endpspicture
\end{center}
\caption{\label{fig:Jdd}
One- and two-baryon processes contained in the used e.m. current. 
In this figure processes are depicted which connect states with a $\Delta$ isobar.
In nonrelativistic order the one-baryon process contributes to charge density
and spatial current, the two-baryon processes only to the spatial current.
Only the  diagonal isovector  $\pi$ exchange is taken into account 
in the two-baryon processes. }
\end{figure*}

The e.m. current acts in a baryonic Hilbert space with two sectors, i.e.,
one sector being purely nucleonic and one in which one nucleon $(N)$ is turned
into a $\Delta$ isobar.
The current operator is employed in its 
Fourier-transformed form $J^{\mu}(\vec{Q})$ and in a momentum representation,
 based on the Jacobi momenta $( \vec{p} \vec{q} \vec{K} )$ of three particles 
in the definition of \Ref~\cite{nemoto:98a}, i.e.,
\begin{gather} \label{eq:JpqK}
\begin{split}
 \langle  \vec{p}' \vec{q}' \vec{K}' | 
J^{\mu} (\vec{Q}) | \vec{p} \vec{q} \vec{K} \rangle = {} & 
\delta (\vec{K}' - \vec{Q} - \vec{K}) \\ &  \times
\langle  \vec{p}' \vec{q}' |j^{\mu} (\vec{Q}, \Kpl) | \vec{p} \vec{q} \rangle .
\end{split}
\end{gather}
In \Eq~\eqref{eq:JpqK} $\vec{Q}$ is the three-momentum transfer by the photon; 
it will take on particular values depending on the considered reaction; 
in the photo reactions of this paper it is given by the photon momentum 
$ \vec{k}_{\gamma}$. A total-momentum conserving $\delta$-function is split off;
 the remaining current operator $j^{\mu} (\vec{Q}, \Kpl)$ 
only acts on the internal momenta of the three-baryon system with a parametric 
dependence on the combination $\Kpl = \vec{K}' + \vec{K}$ of total momenta.
Since all meson degrees of freedom are frozen, the operator has one-baryon 
and many-baryon pieces. Besides the standard nucleonic-current part there 
are additional parts involving the $\Delta$ isobar
which then make effective two- and three-nucleon contributions to the 
exchange current, the contributions being consistent with each other.
We take one-baryon and two-baryon contributions into account, 
shown in \Figs~\ref{fig:Jnn} - \ref{fig:Jdd} and described in detail in the 
respective figure captions. The explicit forms of the considered contributions
are collected in Appendix~\ref{app:current}.
The horizontal lines in the diagrams  indicate that the meson exchanges are 
instantaneous. The dominant meson-exchange contributions  arise
from $\pi$ and $\rho$ exchanges; note, that those are the only contributions
of two-baryon nature taken into account in the calculations of 
\Refs~\cite{golak:00a,skibinski:03a,skibinski:03b}. In our calculations
also the nondiagonal $\rho\pi\gamma$ and $\omega\pi\gamma$ contributions are 
taken into account for the  currents of \Figs~\ref{fig:Jnn} and \ref{fig:Jnd}.
The current of \Fig~\ref{fig:Jnd} couples purely nucleonic states with 
states containing one $\Delta$ isobar.
In contrast to \Ref~\cite{yuan:02a}, the contributions between 
$\Delta$-isobar states of one- and two-baryon nature are kept as shown in 
\Fig~\ref{fig:Jdd}, though the corresponding two-baryon contributions will
turn out to be quantitatively entirely irrelevant; we therefore take only
the diagonal $\pi$ contribution into account.
 The current is derived by the extended $S$-matrix method 
of \Refs~\cite{strueve:87a,adam:89a,adam:91a,henning:92a}; however,
it satisfies current conservation only approximately with the corresponding 
$\pi$ and $\rho$ exchanges in the employed two-baryon interaction $H_I$ of 
CD Bonn and CD Bonn + $\Delta$. 
The spatial current is systematically expanded up to first order in $k/m_N$, 
$k$ being a characteristic baryon momentum and $m_N$ the nucleonic rest mass.
The charge density is used in zeroth order in $k/m_N$ for the standard 
calculations of \Sect~\ref{sec:res}; even photo reactions require the charge
density operator, i.e., for the Siegert form of the current.

In the perturbative spirit for the evolution of photo processes, the e.m. 
interaction $H_I^{\mathrm{e.m.}}$
acts only once, whereas the hadronic interaction $ H_I $ has exactly 
to be taken into account up to all orders. We use hadronic channel states,
seen  in the initial and final states $ | i \vec{P}_i \rangle $ and 
$ | f \vec{P}_f \rangle $ of the photo reactions with total momenta $\vec{P}_i$
and $\vec{P}_f$  in the form
\begin{subequations}  \label{eq:Phi}
  \begin{align}
    | \Phi_B \vec{K} \rangle & = | B \rangle | \vec{K} \rangle ,  \\ 
    | \Phi_{\alpha} (\vec{q}) \nu_{\alpha} \vec{K} \rangle & =  
    | \phi_{\alpha} (\vec{q}) \nu_{\alpha} \rangle | \vec{K} \rangle , \\
    | \Phi_0 (\vec{p} \vec{q}) \nu_0 \vec{K} \rangle & =  
    | \phi_0 (\vec{p} \vec{q}) \nu_0 \rangle | \vec{K} \rangle 
  \end{align} 
\end{subequations}
with the energies
\begin{subequations} \label{eq:E}
  \begin{align}
    E_B (\vec{K}) & = E_B + \frac{\vec{K}^2}{6 m_N}, \\
    E_{\alpha} (\vec{q} \vec{K}) & =  e_d + 
    \frac{3\vec{q} ^2}{4 m_N} + \frac{\vec{K}^2}{6 m_N}, \\
    E_0 (\vec{p} \vec{q} \vec{K}) & =  \frac{\vec{p}^2}{m_N} +
    \frac{3\vec{q} ^2}{4 m_N} + \frac{\vec{K}^2}{6 m_N},
  \end{align} 
\end{subequations}
$m_N$, $e_d$ and $E_B$ being the average rest mass of the nucleon, the deuteron 
and the trinucleon binding energies; in contrast to the notation of
\Ref~\cite{yuan:02a}, but consistent with our notation of hadronic 
reactions~\cite{nemoto:98a,deltuva:03a}, the rest mass of three nucleons
is removed from the energies of \Eqs~\eqref{eq:E}.
 The internal trinucleon bound state is $|B \rangle$, 
which is normalized to 1. The product nucleon-deuteron and breakup channel states 
in the three-nucleon c.m. frame are $|\phi_{\alpha} (\vec{q}) \nu_{\alpha}\rangle$
 and $|\phi_0 (\vec{p} \vec{q}) \nu_0 \rangle $ in the notation of 
\Ref~\cite{nemoto:98a}, $\nu_{\alpha}$ and $\nu_0$ denoting all discrete 
quantum numbers. In both cases the c.m. motion $|\vec{K} \rangle $
is explicitly added to the 
internal motion; in the three-nucleon channel with a photon, the total momentum 
$\vec{P}$ is different from the total momentum $\vec{K}$ of 
the three nucleons, bound in the trinucleon bound state $|B \rangle$;
in the channels without photon $\vec{P} = \vec{K}$.

The matrix elements of the e.m. interaction require fully correlated 
hadronic states, i.e.,
\begin{subequations} \label{eq:Psi}
  \begin{align} \label{eq:Psia}
    | \Phi_B \vec{K} \rangle  = &\pm i0 \, G \big(E_B (\vec{K}) \pm i0 \big) 
    | B \rangle | \vec{K} \rangle, \\  
    \label{eq:Psib}
    | \Psi^{(\pm)}_{\alpha} (\vec{q}) \nu_{\alpha} \vec{K} \rangle  = &  
    \pm i0 \, G \big(E_{\alpha} (\vec{q} \vec{K}) \pm i0 \big)  
    \nonumber \\ & \times
    { \frac{1}{\sqrt{3}} } \big( 1+P \big)  
    |\Phi_{\alpha} (\vec{q}) \nu_{\alpha} \vec{K} \rangle, \\
    \label{eq:Psic}
    | \Psi^{(\pm)}_0 (\vec{p} \vec{q}) \nu_0 \vec{K} \rangle =  &
    \pm i0 \, G \big(E_0 (\vec{p} \vec{q} \vec{K}) \pm i0 \big) 
    \nonumber \\ & \times
    { \frac{1}{\sqrt{3}} } \big( 1+P \big) 
    |\Phi_0 (\vec{p} \vec{q}) \nu_0 \vec{K} \rangle, 
  \end{align} 
\end{subequations}
with the full resolvent
\begin{gather}
  G(Z) = \big( Z - H_0 - H_I \big)^{-1} ;
\end{gather}
the free Hamiltonian $ H_0 $ contains the motion of the center of mass,
but the rest mass of three nucleons is taken out, consistent with 
\Eqs~\eqref{eq:E}; 
the permutation operator $P$ symmetrizes the product states; 
the individual kinetic energy operators are of nonrelativistic form; 
they yield the eigenvalues of \Eqs~\eqref{eq:E}. The hadronic states 
\eqref{eq:Psib} and \eqref{eq:Psic} are normalized to $\delta$
functions without additional normalization factors. Since the hadronic 
interaction Hamiltonian $H_I$ acts on relative coordinates only, 
the full resolvent reproduces the bound state $|B \rangle $ and correlates 
the scattering states only in their internal parts, i.e.,
\begin{subequations} \label{eq:psi}
  \begin{align}    \label{eq:psia}
    | \Psi^{(\pm)}_{\alpha} (\vec{q}) \nu_{\alpha} \vec{K} \rangle & = 
    | \psi^{(\pm)}_{\alpha} (\vec{q}) \nu_{\alpha} \rangle |\vec{K} \rangle, \\ 
    \label{eq:psib}
    | \Psi^{(\pm)}_0 (\vec{p} \vec{q}) \nu_0 \vec{K} \rangle & =  
    | \psi^{(\pm)}_0 (\vec{p} \vec{q}) \nu_0 \rangle |\vec{K} \rangle .
  \end{align} 
\end{subequations}

\subsection{ \label{sec:smat} $S$ matrix for three-body photo disintegration
of the trinucleon bound state}

\begin{figure}[!b]
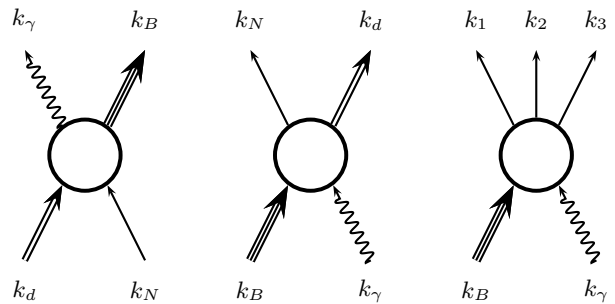

\begin{center}
\pspicture(0,0)(8.0,4.0)
\def\nucleon{\psline{->}(0,0)(-0.5,1.0)}
\def\trinucl{\psline[doubleline=true,doublesep=2pt]{->}(0,0)(0.5,1.0)
\psline(0,0)(0.45,0.9)}
\def\deutr{\psline[doubleline=true,doublesep=1pt]{->}(0,0)(0.5,1.0)}
\def\photon{\pscoil[coilwidth=0.15cm,coilaspect=0,coilarmA=0cm,coilarmB=0.2cm]
{->}(0,0)(-0.5,1.0)}
\multips(1,2)(3,0){3}{\pscircle[linewidth=1.5pt]{0.5cm}}
\rput(0.2,0.6){\deutr}   \rput(0.2,0.2){$k_d$}
\rput(1.8,0.6){\nucleon} \rput(1.8,0.2){$k_N$}
\rput(1.3,2.4){\trinucl} \rput(1.8,3.8){$k_B$}
\rput(0.7,2.4){\photon}  \rput(0.2,3.8){$k_{\gamma}$}
\multips(3.2,0.6)(3,0){2}{\trinucl} \multirput(3.2,0.2)(3,0){2}{$k_B$}
\multips(4.8,0.6)(3,0){2}{\photon}  \multirput(4.8,0.2)(3,0){2}{$k_{\gamma}$}
\multips(3.7,2.4)(3,0){2}{\nucleon} \rput(3.2,3.8){$k_N$} \rput(6.2,3.8){$k_1$} 
\rput(4.3,2.4){\deutr}              \rput(4.8,3.8){$k_d$}
\psline{->}(7,2.5)(7,3.4)           \rput(7.0,3.8){$k_2$}
\psline{->}(7.3,2.4)(7.8,3.4)       \rput(7.8,3.8){$k_3$}
\endpspicture
\end{center}
\caption{\label{fig:reaction}
Schematic description of all considered three-nucleon photo reactions. 
The lines for the two-baryon and three-baryon particles are drawn in a special 
form to indicate their compositeness.}
\end{figure}

The $S$-matrix and the spin-averaged and spin-dependent cross sections for
radiative nucleon-deuteron capture and for two-body photo disintegration 
of the trinucleon bound  state are given in \Ref~\cite{yuan:02a}. 
We add now the corresponding quantities for three-body photo disintegration.
The kinematics of all considered photo processes is shown in 
\Fig~\ref{fig:reaction}.
The figure also defines the employed notation for the individual particle momenta
of the trinucleon bound state, deuteron, a nucleon, the three-nucleons of break-up
and the photon; 
 $k_B$, $k_d$, $k_N$, $k_i$ and $k_{\gamma}$ are on-mass-shell four-momenta.
 The corresponding particle energies are the zero components of those momenta, 
i.e.,  $k_B^0 c$, $k_d^0 c$, $k_N^0 c$, $k_i^0 c$ and $k_{\gamma}^0 c$; 
they are relativistic ones with the complete rest masses
in contrast to those of the nonrelativistic model calculation of baryonic 
states in \Eqs~\eqref{eq:E}.

We give various alternative forms for the $S$-matrix elements:
\begin{widetext}
\begin{subequations} \label{eq:Smat}
  \begin{align} \label{eq:Smata}
    \langle f \vec{P}_f | S | i \vec{P}_i \rangle  = &  
    -i(2 \pi \hbar)^{4} \delta ( k_{1} + k_{2} + k_{3} - k_{\gamma} - k_B)
    \, \langle s_f | M | s_i \rangle    \,
    (2 \pi \hbar)^{-15/2} \big[ 2 k_{\gamma}^{0}c \, 2 k_B^0 c \, 
      2k_1^0 c \, 2k_2^0 c \, 2k_3^0 c \big]^{-1/2},  \\
    \label{eq:Smatb}
    \langle f \vec{P}_f | S | i \vec{P}_i \rangle  = &  
    -2 \pi i \, \delta \big( E_N(\vec{k}_{1}) + E_N(\vec{k}_{2}) +
    E_N(\vec{k}_{3}) - k_{\gamma}^{0}c - E_B(\vec{k}_B) \big)  \,
    \langle f \vec{P}_f | i0 G(E_i+i0) H^{e.m.}_I | i \vec{P}_i\rangle, \\
    \label{eq:Smatc}
    \langle f \vec{P}_f | S | i \vec{P}_i \rangle  = &  
    -2 \pi i \, \delta \big( E_N(\vec{k}_{1}) + E_N(\vec{k}_{2}) +
    E_N(\vec{k}_{3}) - k_{\gamma}^{0}c - E_B(\vec{k}_B) \big)  \,
    \delta (\vec{k}_{1} + \vec{k}_{2} + \vec{k}_{3} - 
    \vec{k}_{\gamma} - \vec{k}_B ) 
    \nonumber \\     & \times    
    \frac{(4\pi)^{1/2} \hbar }{(2\pi\hbar)^{3/2} (2k_{\gamma}^{0}c)^{1/2}} \,
    \langle \psi^{(-)}_0 (\vec{p}_f \vec{q}_f) \nu_{0f} |
    j^{\mu} (\vec{k}_{\gamma}, \Kpl) 
    \epsilon_{\mu} (\vec{k}_{\gamma} \lambda) | B \rangle. 
  \end{align}
\end{subequations}
\end{widetext}
Equation~\eqref{eq:Smata} introduces a covariant form, whereas 
\Eqs~\eqref{eq:Smatb} and \eqref{eq:Smatc} are noncovariant quantum mechanical
realizations of it.
$ \epsilon_{\mu} (\vec{k}_{\gamma} \lambda) $ is the polarization vector 
of the real photon with helicity $\lambda$.
$ \langle s_f | M | s_i \rangle $ is the singularity-free matrix element for 
three-nucleon photo disintegration, from which the differential cross section
\begin{gather} \label{eq:dsigma}
  d \sigma_{i \to f}  =  \big | \langle s_f | M  | s_i \rangle \big |^2 
  \frac{d\mathrm{Lips} (k_{\gamma}+k_B, k_{1}, k_{2}, k_{3})}
       {4c^2 \, k_B \cdot k_{\gamma}} 
\end{gather}
 is obtained. Its dependence on the helicity $\lambda $ of the photon and on 
the spin projection ${\mathcal M}_{B}$ of the trinucleon bound state 
in the initial channel,
 collectively described by $s_i$, and on the spin projections $m_{s_f}$ of 
nucleons  in the final channel, collectively described by $s_f$, are explicitly 
indicated. $\langle s_f | M | s_i \rangle $ is Lorentz-invariant  in a 
relativistic description and can therefore be calculated in any frame. 
However, in our model it is calculated in the framework of nonrelativistic 
quantum mechanics and therefore loses the property of being a Lorentz scalar;
equating \Eqs~\eqref{eq:Smata} and \eqref{eq:Smatc} 
$\langle s_f | M | s_i \rangle $  is  defined by 
\begin{gather} \label{eq:Mampl}
  \begin{align} 
    \langle s_f | M | s_i \rangle = & \frac{\sqrt{4\pi}}{c} (2 \pi \hbar)^{3}
    \big[ 2k_B^0 c \, 2k_1^0 c \, 2k_2^0 c \, 2k_3^0 c \big]^{1/2} \nonumber  \\
    &  \times \langle \psi^{(-)}_0 (\vec{p}_f \vec{q}_f) \nu_{0f} |
    j^{\mu} (\vec{k}_{\gamma}, \Kpl)
    \epsilon_{\mu} (\vec{k}_{\gamma} \lambda) | B \rangle.
  \end{align}
\end{gather}
We calculate that matrix element  in the center of mass (c.m.) system of the 
final hadronic state using the following computational strategy. 
The strategy is nonunique, since the model calculations,
 due to dynamic limitations, miss the trinucleon binding energy;
the necessary correction for that miss has arbitrary features.
In contrast, the $S$ matrix of \Eq~\eqref{eq:Smata} is based
on proper relativistic kinematics with experimental rest masses. 

1. The experimental photon momentum $\kgl$ in the lab   frame with 
  $\vec{k}_B=0$ determines the total momentum $\vec{P}_f$ and energy
  $E_0 (\vec{p}_f \vec{q}_f \vec{K}_f)$ of the final three-nucleon system
  in the lab frame, 
  i.e., $\vec{P}_f = \vec{K}_f = \kgl $ and 
  $E_0 (\vec{p}_f \vec{q}_f \vec{K}_f) = E_B + |\kgl| c$.
  This step is done using the experimental trinucleon  binding energy.
  The resulting energy $E_0 (\vec{p}_f \vec{q}_f \vec{K}_f)$ of the final state
  is the true experimental one. Thus, the experimental 
  two-body and three-body breakup thresholds are exactly reproduced.
  In nonrelativistic approximation
  for baryon kinematics, the internal three-nucleon kinetic energy part
  of the final state is  ${\vec{p}_f^2}/{m_N} + {3\vec{q}_f^2}/{4 m_N} = 
  E_0 (\vec{p}_f \vec{q}_f \vec{K}_f)  - {\vec{K}_f^2}/{6 m_N}$.

2. The matrix element $\langle s_f | M | s_i \rangle $ is calculated 
  in the c.m. system as \emph{on-energy-shell element}
  under nonrelativistic model assumptions. 
  Under those assumptions the internal energy of the initial state is
  $|\kgc|c + E_B + \kgc^2/6m_N = 
  E_0 (\vec{p}_f \vec{q}_f \vec{K}_f)  - {\vec{K}_f^2}/{6 m_N}$,
  $\kgc$ being the photon momentum in the c.m. system,
  in which the trinucleon bound state is moving with momentum 
  $\vec{k}_B = -\kgc$; thus, $\Kpl = -\kgc$.
  Taking the computed trinucleon model binding energy $E_B$
  and the average nucleon mass $m_N $, i.e., 
  $m_N c^2 = 938.919$~MeV, the magnitude of the photon momentum 
  $|\kgc|$ to be used for the current matrix element results.
  Since the model binding energy $E_B $ is   not the experimental one, 
  neither for ${}^3\mathrm{He}$ nor for ${}^3\mathrm{H}$, 
  and since  the c.m. contribution to total three-nucleon energies is assumed
  to be nonrelativistic with mass $3m_N$ and to separate from its 
  internal part, that photon momentum $\kgc$ does not have the 
  experimental value.  In  the matrix element $\langle s_f | M | s_i \rangle$,
  to be calculated according to \Eq~\eqref{eq:Mampl},
  the zero-momentum components $k_i^0$ and $k_B^0$ are nonrelativistic model
  quantities and differ from the baryonic energies just by rest masses, i.e.,
  $k_i^0 c = E_N(\vec{k}_i) + m_N c^2$ and $k_B^0 c = E_B(\vec{k}_B) + 3m_N c^2$.

In contrast to the matrix element $\langle s_f | M | s_i \rangle $
which carries the dynamics, the kinematical factors in \Eq~\eqref{eq:dsigma},
i.e., the Lorentz-invariant phase-space element
\begin{gather} \label{eq:dLips}
  \begin{split}
    d \mathrm{Lips}  (k_{\gamma} \!+\! k_B, & k_{1}, k_{2}, k_{3}) 
    \\ 
    = {} &   ( 2\pi \hbar)^4 \, \delta 
    ( k_{1} + k_{2} + k_{3} - k_{\gamma} - k_B) \\
    & \times  \frac{d^3 k_{1} \, d^3 k_{2} \, d^3 k_{3}}
    {(2\pi \hbar)^9 \, 2k^0_{1}c \, 2k^0_{2}c \, 2k^0_{3}c}
  \end{split}
\end{gather}
and the factor $4c^2\, k_B \cdot k_{\gamma}$,
which contains the incoming flux, the target density and projectile and target 
normalization factors can be calculated relativistically.

The  momenta in the initial and final states are constrained by energy and
momentum conservation. E.g., if  the momentum $\vec{k}_{1}$ and
the direction $\hat{\vec{k}}_{2}$ were measured, all three nucleon momenta are
determined in the final state, although not always uniquely.
In practice, the two nucleon scattering angles with respect to the beam direction
$(\theta_1, \varphi_1)$ and $(\theta_2, \varphi_2)$,
usually notationally shortened to 
$(\theta_1, \theta_2, \varphi_2 - \varphi_1)$,
and their kinetic energies without rest masses $E_{1} = E_N(\vec{k}_1)$ 
and $E_{2} =E_N(\vec{k}_2) $ are measured. Those energies 
are related by momentum and energy conservation and therefore lie on a fixed
kinematical curve. The observables are therefore given as function of the
arclength $S$ along that curve, i.e.,
\begin{gather} \label{eq:Scurve}
  S  = \int _0 ^S d S
\end{gather}
with $dS = \sqrt{ d E_{1}^2 + d E_{2}^2}$ and $E_{2}$ 
being considered a function of $E_{1}$ or vice versa depending on
numerical convenience; the arclength is always taken counterclockwise
along the kinematical curve.
No  confusion between the arclength $S$ and the $S$-matrix of \Eq~\eqref{eq:Smat} 
should arise. The normalization of the arc length value zero is chosen 
as $\max \{ E_{1} | E_{2} = 0 \}$.

The lab cross section therefore takes the compact form
\begin{subequations} \label{eq:d5Sr}
  \begin{gather}
    \begin{split} \label{eq:d5s}
      d^5 \sigma_{i \to f} = & 
       \big | \langle s_f | M | s_i \rangle \big |^2
       \mathrm{fps} \: d S \, d^2 \mhk_{1} \, d^2 \mhk_{2}
    \end{split}
\end{gather}
with the abbreviation $\mathrm{fps}$  for a phase-space factor; 
in the lab frame $\mathrm{fps}$ is
\begin{gather} \label{eq:fps}
  \begin{align} 
    \mathrm{fps} = & \frac{(2 \pi \hbar)^{-5}}{4 c^3 k_{\gamma}^0 m_B} 
    \int d^3  k_{3} \, \vec{k}^2_{2} \, d k_{2} 
    \left( \frac {\vec{k}^2_{1} d k_{1}} {d S} \right) \nonumber  \\
    & \times \frac{\delta (k_{1}  + k_{2} + k_{3} - k_{\gamma} - k_B) }
    {2k^0_{1}c \, 2k^0_{2}c \, 2k^0_{3}c}, \\
    \mathrm{fps} = & \frac{(2 \pi \hbar)^{-5}}{32 c^7 k_{\gamma}^0 m_B }
    \, \vec{k}_1^2  \vec{k}_2^2   \nonumber \\
    & \times \Big \{ \vec{k}_1^2   \big[ |\vec{k}_2| (k_2^0 + k_3^0) 
      - k_2^0 \mhk_2 \cdot (\kgl  \! - \! \vec{k}_1) \big]^2  \nonumber \\
    & + \vec{k}_2^2 \big[ |\vec{k}_1| (k_1^0+k_3^0)  - 
      k_1^0 \mhk_1 \cdot (\kgl  \! - \! \vec{k}_2) \big]^2 \Big \} ^{-1/2}.
  \end{align}
\end{gather}
\end{subequations}
  The cross section~\eqref{eq:d5s} is still spin-dependent.
  The spin-averaged fivefold differential cross section is
\begin{gather} \label{eq:d5s-av}
  \frac{d^5 \sigma}{dS \, d\Omega_1 \, d\Omega_2} = 
     \frac 14 \sum_{\mathcal{M}_B \lambda} \: \sum _{m_{s_1} m_{s_2} m_{s_3}} 
      \frac {d^5 \sigma_{i\to f}} {d S \, d^2 \mhk _{1} \, d^2 \mhk_{2} } .
\end{gather}
Spin observables are defined as in \Refs~\cite{nemoto:98b,chmielewski:03a}.
The experimental setup determines the isospin character of 
the two detected nucleons 1 and 2.

The calculational strategy of \Eqs~\eqref{eq:Mampl} - \eqref{eq:d5Sr} is in
the spirit of \Ref~\cite{yuan:02a}; it chooses the kinematics differently
for the dynamic matrix element $\langle s_f | M | s_i \rangle $ on one side
and for the phase space 
$d \mathrm{Lips}  (k_{\gamma} \!+\! k_B, k_{1}, k_{2}, k_{3}) $ and the 
factor $ 4c^2 k_B \cdot k_{\gamma}$ on the other side. That strategy 
can be carried out with ease for the observables of
exclusive processes. However, when total cross sections in hadronic and 
e.m. reactions or inelastic structure functions in electron scattering 
are calculated as described in Appendix~\ref{sec:inteq} for the total
photo cross section, the energy conserving
 $\delta$ function is rewritten as imaginary part of the full resolvent
and has to be made consistent with the employed nonrelativistic dynamics.
Thus, as described in Appendix~\ref{sec:inteq}, the  split calculational 
strategy, developed in \Ref~\cite{yuan:02a} and so far here,
cannot be carried through for total cross sections and inelastic structure 
functions; furthermore, as discussed in \Ref~\cite{chmielewski:03a}
for the hadronic reactions, such a split calculational strategy would also be
inconsistent with the fit of the underlying baryonic potentials.

We shall therefore use nonrelativistic kinematics in the framework of 
quantum mechanics throughout. 
The corresponding expressions, derived from  quantum mechanics directly,
can also be obtained formally from \Eqs~\eqref{eq:Mampl} - \eqref{eq:dLips}
based on quantum field theory by  
replacing the hadron energy factors $2k_j^0 c$ by their rest masses $2m_j c^2$
and using nonrelativistic energies for the energy conserving $\delta$-functions
and for the definition of the kinematic locus.
The lab cross section is constructed from the following building blocks, 
i.e., the matrix element
\begin{subequations} \label{eq:d5Sn}
\begin{gather} \label{eq:MamplN}
  \begin{split} 
    \langle s_f | M | s_i \rangle = & \frac{\sqrt{4\pi}}{c} (2 \pi \hbar)^{3}
    \big[ 2m_B c^2 \, (2m_N c^2)^3 \big]^{1/2}   \\
    &  \times \langle \psi^{(-)}_0 (\vec{p}_f \vec{q}_f) \nu_{0f} |
    j^{\mu} (\kgc, -\kgc)  \\ & \times
    \epsilon_{\mu} (\kgc \lambda) | B \rangle
  \end{split}
\end{gather}
and the phase space factor $\mathrm{fps}$ of \Eqs~\eqref{eq:d5Sr}
which takes the following changed form
\begin{gather} \label{eq:fpsn}
  \begin{align} 
    \mathrm{fps} = & \frac{(2 \pi \hbar)^{-5}} {4 c^2 k_{\gamma}^0 m_B } 
     \int d^3  k_{3} \, \vec{k}^2_{2} \, d k_{2} 
    \left( \frac {\vec{k}^2_{1} d k_{1}} {d S} \right)   \nonumber \\
    & \times \frac{ \delta (E_N(\vec{k}_{1}) + E_N(\vec{k}_{2}) +
    E_N(\vec{k}_{3}) - k_{\gamma}^{0}c - E_B ) } 
    {(2m_N c^2)^{3}}  \nonumber \\
    & \times \delta (\vec{k}_{1} + \vec{k}_{2} + \vec{k}_{3} - \kgl), \\
    \mathrm{fps} = & \frac{(2 \pi \hbar)^{-5}} {32 c^8 k_{\gamma}^0 m_N m_B }
    \, \vec{k}_1^2  \vec{k}_2^2   
    \Big \{ \vec{k}_1^2 \big[ 2|\vec{k}_2|  - 
      \mhk_2 \cdot (\kgl  \! - \! \vec{k}_1) \big]^2  \nonumber \\
    & + \vec{k}_2^2 \big[ 2|\vec{k}_1|  - 
      \mhk_1 \cdot (\kgl  \! - \! \vec{k}_2) \big]^2 \Big \} ^{-1/2}.
  \end{align}
\end{gather}
\end{subequations}
Section~\ref{sec:shortcomings} will discuss the differences between the present
fully nonrelativistic calculational scheme of cross sections and that
of \Eqs~\eqref{eq:Mampl} and \eqref{eq:d5Sr} with some relativistic features.

\section{ \label{sec:res} Results}

We present results for spin-averaged and spin-dependent observables 
of nucleon-deuteron radiative capture and of three-nucleon photo disintegration;
results of two-nucleon photo disintegration are transformed to corresponding
ones of radiative capture.
The results are based on calculations derived from the purely nucleonic
CD-Bonn potential~\cite{machleidt:01a} and its coupled-channel 
extension~\cite{deltuva:03c}, which allows for single $\Delta$-isobar
excitation in isospin-triplet partial waves. 
The $\Delta$ isobar is considered to be a stable particle of spin and 
isospin 3/2 with a rest mass $m_{\Delta}c^2$ of 1232~MeV.
In contrast to the coupled-channel potential constructed previously by the 
subtraction technique~\cite{hajduk:83a} and used in the calculations of
\Ref~\cite{yuan:02a}, the new one of \Ref~\cite{deltuva:03c} is fitted properly 
to data and accounts for two-nucleon scattering data with the same quality as the
original CD-Bonn potential. We describe first the \emph{standard calculational 
procedure} adopting the strategy of \Sect~\ref{sec:smat}.

The hadronic interaction in purely nucleonic and in nucleon-$\Delta$ partial 
waves up to the total two-baryon angular momentum $I=4$ is taken into account.
The calculations omit the Coulomb potential between charged baryons.
Nevertheless, the theoretical description is charge dependent. For reactions on
${}^3\mathrm{He}$ the $pp$ and $np$ parts of the interaction are used, 
for reactions on ${}^3\mathrm{H}$ the $nn$ and $np$ parts.
Assuming charge independence, the trinucleon bound state and
nucleon-deuteron scattering states are pure states with total isospin
$\mathcal{T} = \frac12$; the three-nucleon scattering states have total
isospin $\mathcal{T} = \frac12$ and $\mathcal{T} = \frac32$, but those
parts are not dynamically coupled.  Allowing for charge dependence,
all three-baryon states have $\mathcal{T} = \frac12$ and $\mathcal{T} = \frac32$
components which are dynamically coupled. 
For hadronic reactions that coupling is found to be quantitatively important 
in the ${}^1S_0$ partial wave~\cite{deltuva:03b}; in other partial waves
the approximative treatment of charge dependence as described in 
\Ref~\cite{deltuva:03b} is found to be sufficient; it does not  couple
total isospin $\mathcal{T} = \frac12$ and $\frac32$ channels dynamically.
The same applies for photo reactions considered in this paper. 
The effect of charge dependence is dominated by the ${}^1S_0$ partial wave; 
it is seen in some particular kinematics of radiative capture
and of three-body photo disintegration; we do not discuss it in this paper.
Furthermore, the calculations of e.m. reactions require  total isospin
$\mathcal{T} = \frac32$ components of scattering states
in \emph{all} considered isospin-triplet
two-baryon partial waves, since the e.m. current couples the 
$\mathcal{T} = \frac12$ and $\mathcal{T} = \frac32$ components strongly.

The  three-particle equations for the trinucleon bound state $|B \rangle$ 
and for the scattering states are solved as in \Ref~\cite{deltuva:03a};
in fact, the scattering states are calculated only implicitly as described
in Appendix~\ref{sec:inteq}. The resulting binding energies
of ${}^3\mathrm{He}$ are -7.941 and -8.225~MeV for CD Bonn and CD Bonn + $\Delta$,
respectively.  If the Coulomb interaction were taken into account, 
as proper for ${}^3\mathrm{He}$, the binding energies shift to
 -7.261 and -7.544~MeV, whereas the experimental value is -7.718~MeV.
  Nevertheless, we use the purely hadronic energy values
and bound-state wave functions for consistency when calculating the current 
matrix elements, since we are unable to include the Coulomb interaction 
in the scattering states.

Whereas the hadronic interaction is considered up to $I=4$, the e.m.
 current is allowed to act between partial waves up to $I=6$,
the higher partial waves being created by the geometry of antisymmetrization.
The e.m. current is taken over from \Refs~\cite{strueve:87a,oelsner:phd}
with some necessary modifications: 1) The e.m. current is richer than the one
used in \Ref~\cite{yuan:02a}; diagonal two-baryon currents connecting states
with $\Delta$ isobar are taken into account.
2) More recent values for the  e.m. couplings of the $\Delta$ isobar 
are used according to \Refs~\cite{carlson:86a,lin:91a}.
3) Meson coupling constants, meson masses and hadronic form factors used
in meson-exchange currents (MEC) are chosen consistently with the employed 
hadronic interactions CD Bonn and CD Bonn + $\Delta$; they are listed
in \Refs~\cite{machleidt:01a,deltuva:03c}. The employed contributions to the 
e.m. current are collected in Appendix~\ref{app:current}.
The current is expanded in electric and magnetic multipoles as described
in \Refs~\cite{yuan:02a,oelsner:phd}. The technique for  calculating multipole
 matrix elements is developed in \Ref~\cite{oelsner:phd}; a special
 stability problem~\cite{yuan:02a} arising in the calculation requires some 
modifications of that technique as described in \Ref~\cite{deltuva:phd}.
The magnetic multipoles are calculated 
 from the one- and two-baryon parts of the spatial current.
The electric multipoles use the Siegert form of the current \emph{without}
long-wavelength approximation;  assuming  current conservation,
the dominant parts of the one-baryon convection
current and of the diagonal $\pi$- and $\rho$-exchange current are taken
into account implicitly in the Siegert part of the electric multipoles
by the Coulomb multipoles of the charge density;
the remaining non-Siegert part of the electric multipoles not
 accounted for by the charge density is calculated using explicit
one- and two-baryon spatial currents.  The charge density contributing
to the Siegert term has diagonal single-nucleon and single-$\Delta$ isobar
contributions only; the nucleon-$\Delta$ transition contribution as well as 
two-baryon contributions are of relativistic order and are therefore omitted 
in the charge-density operator when calculating Coulomb multipoles. 

The number of considered current multipoles is limited by the maximal total 
three-baryon angular momentum $\mathcal{J}_{\mathrm{max}} = \frac{15}{2}$,
taken into account for the hadronic scattering states.
The results for the considered photo reactions up to pion-production threshold
appear fully converged with respect to higher 
two-baryon angular momenta $I$, with respect to $\Delta$-isobar coupling and
with respect to higher three-baryon angular momenta $\mathcal{J}$
on the scale of accuracy which present-day experimental data require.

That is the \emph{ standard calculational procedure}.
 Section~\ref{sec:shortcomings} describes the shortcomings of 
that standard description.  In the rest of this section  we focus on 
$\Delta$-isobar effects in sample observables.

\subsection{\label{sec:rc} Nucleon-deuteron radiative capture}

\begin{figure*}[!]
\includegraphics[scale=\scl]{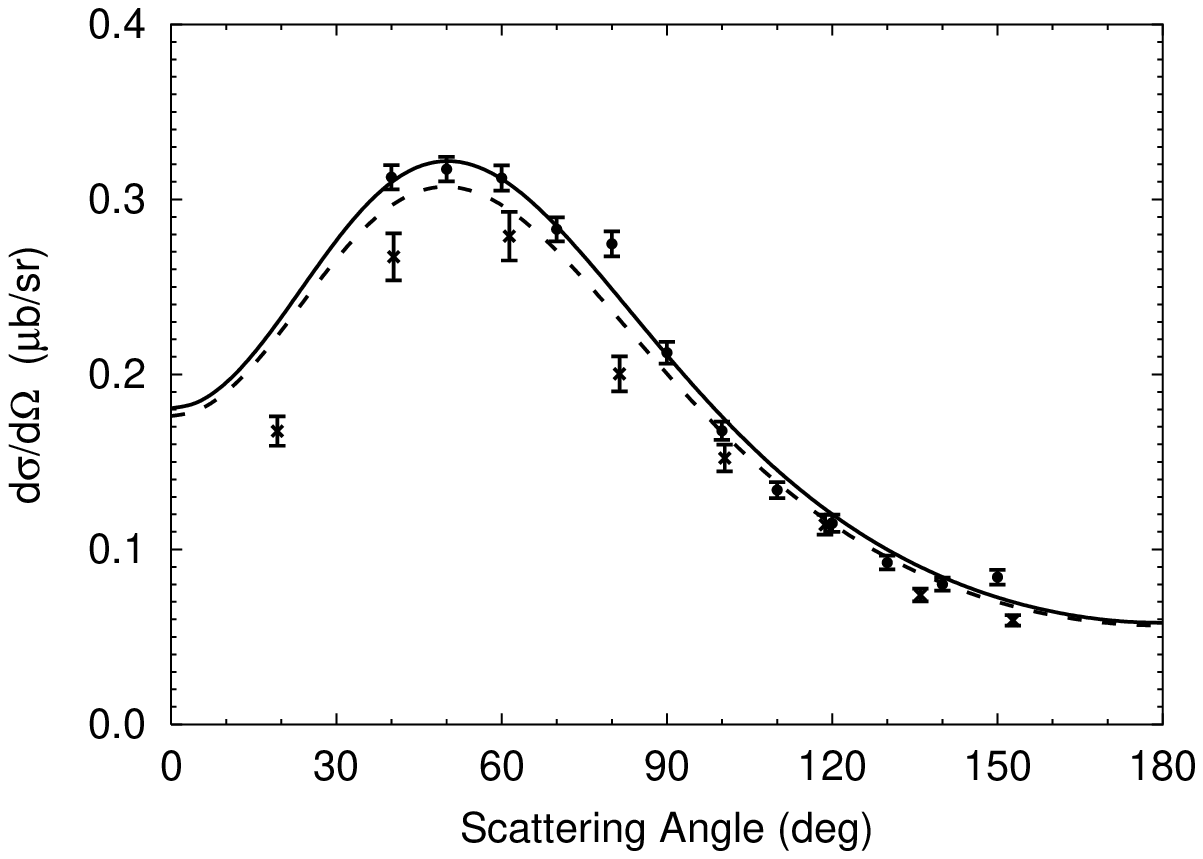} \hspace{5mm}
\includegraphics[scale=\scl]{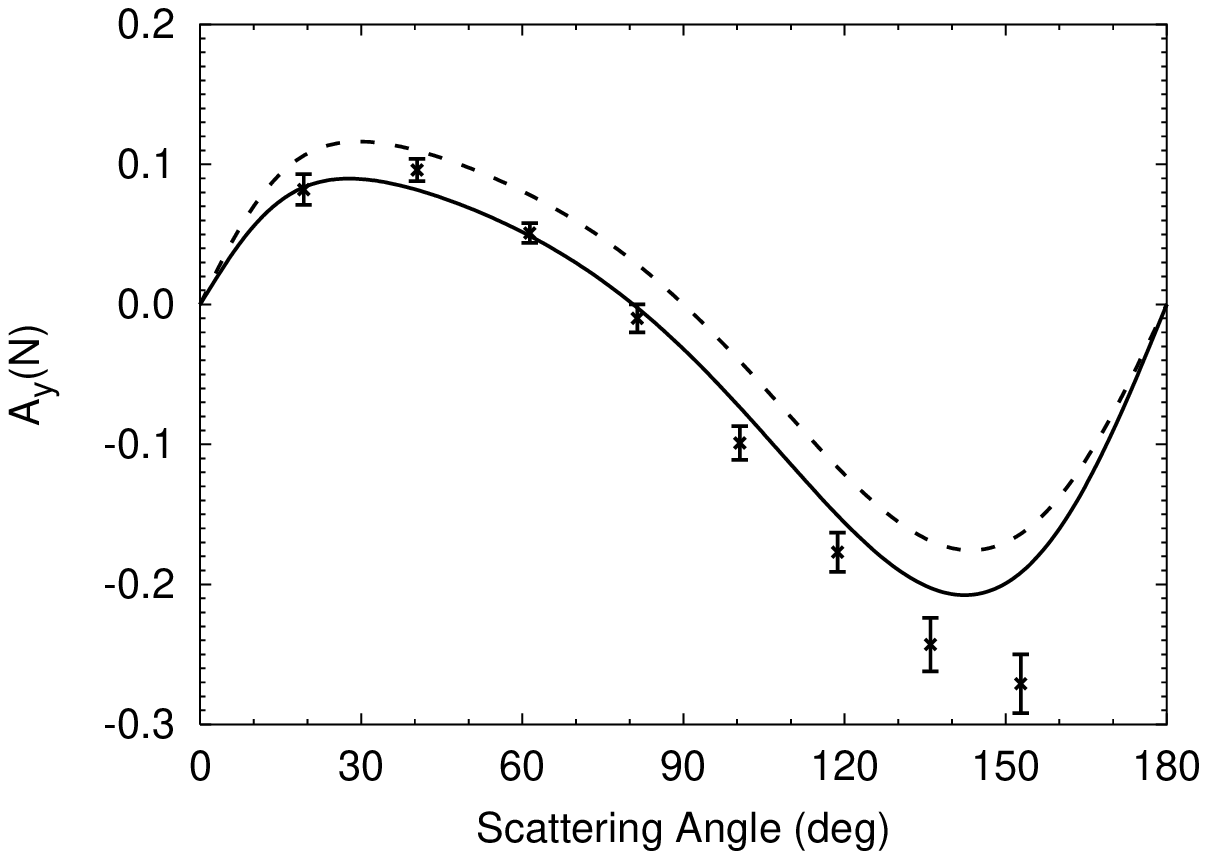} \\  \vspace{1mm}
\includegraphics[scale=\scl]{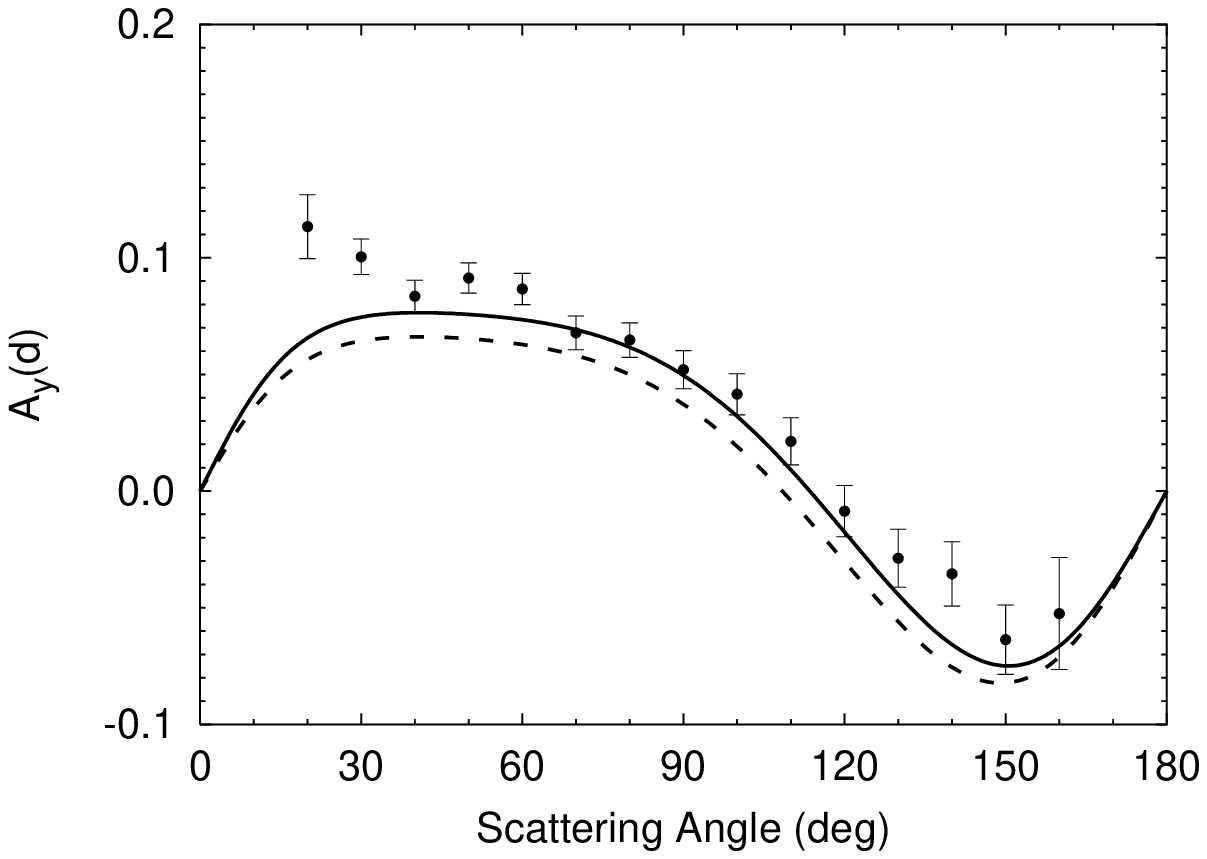} \hspace{5mm}
\includegraphics[scale=\scl]{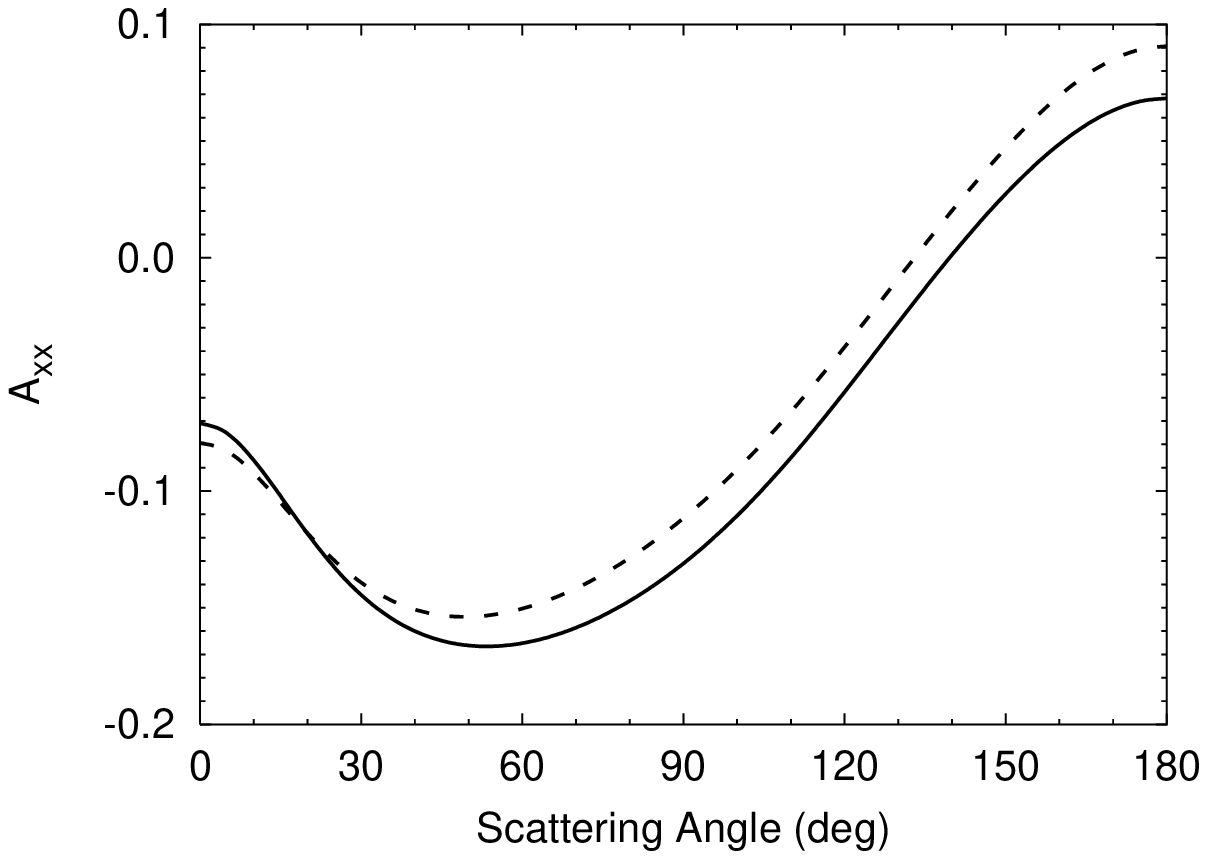} 
\caption{\label{fig:rc100}
Differential cross section and analyzing powers of proton-deuteron 
radiative capture at 100~MeV  nucleon lab energy as function of the c.m. 
nucleon-photon scattering angle.
Results of the coupled-channel potential with $\Delta$-isobar excitation 
(solid curves) are compared with reference results of the purely nucleonic 
CD-Bonn potential (dashed curves).
The experimental data are from \Ref~\cite{yagita:03a} (circles) and
from \Ref~\cite{pickar:87a} (crosses).}
\end{figure*}

\begin{figure*}[!]
\includegraphics[scale=\scl]{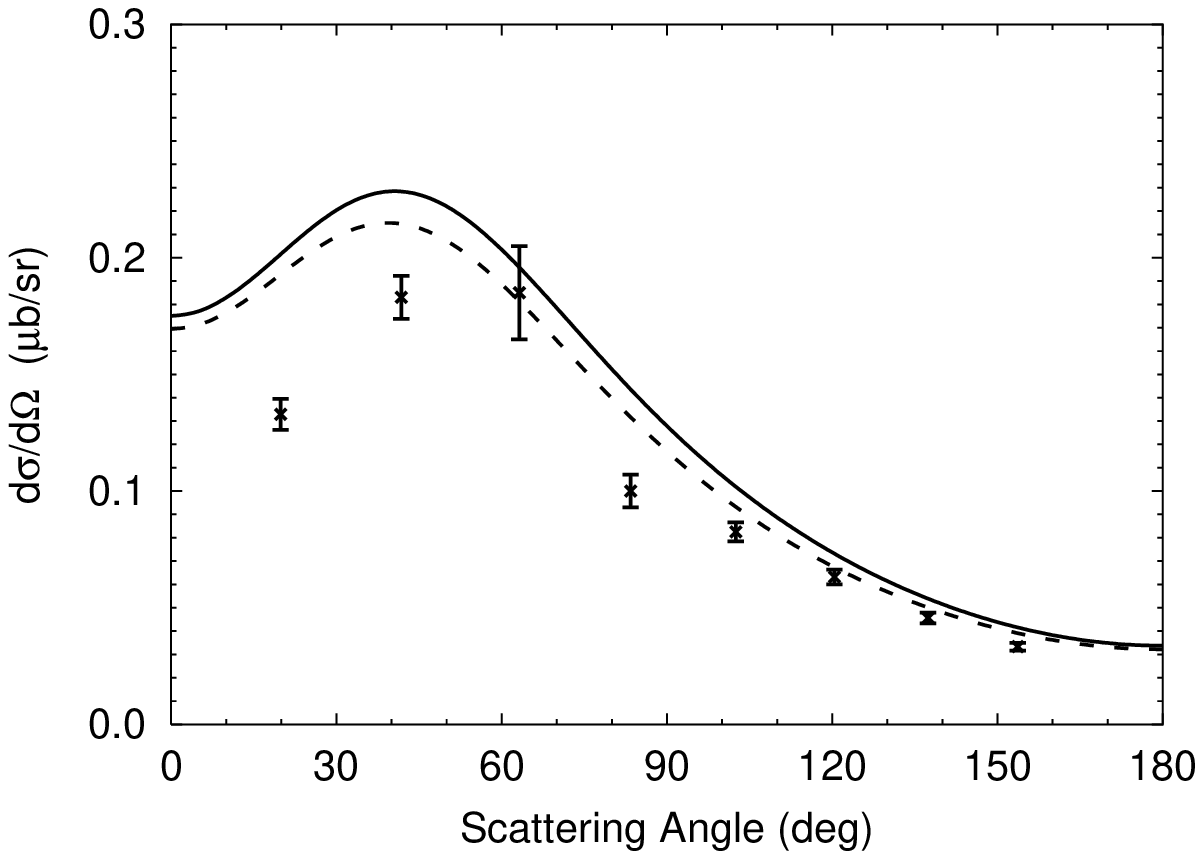} \hspace{5mm}
\includegraphics[scale=\scl]{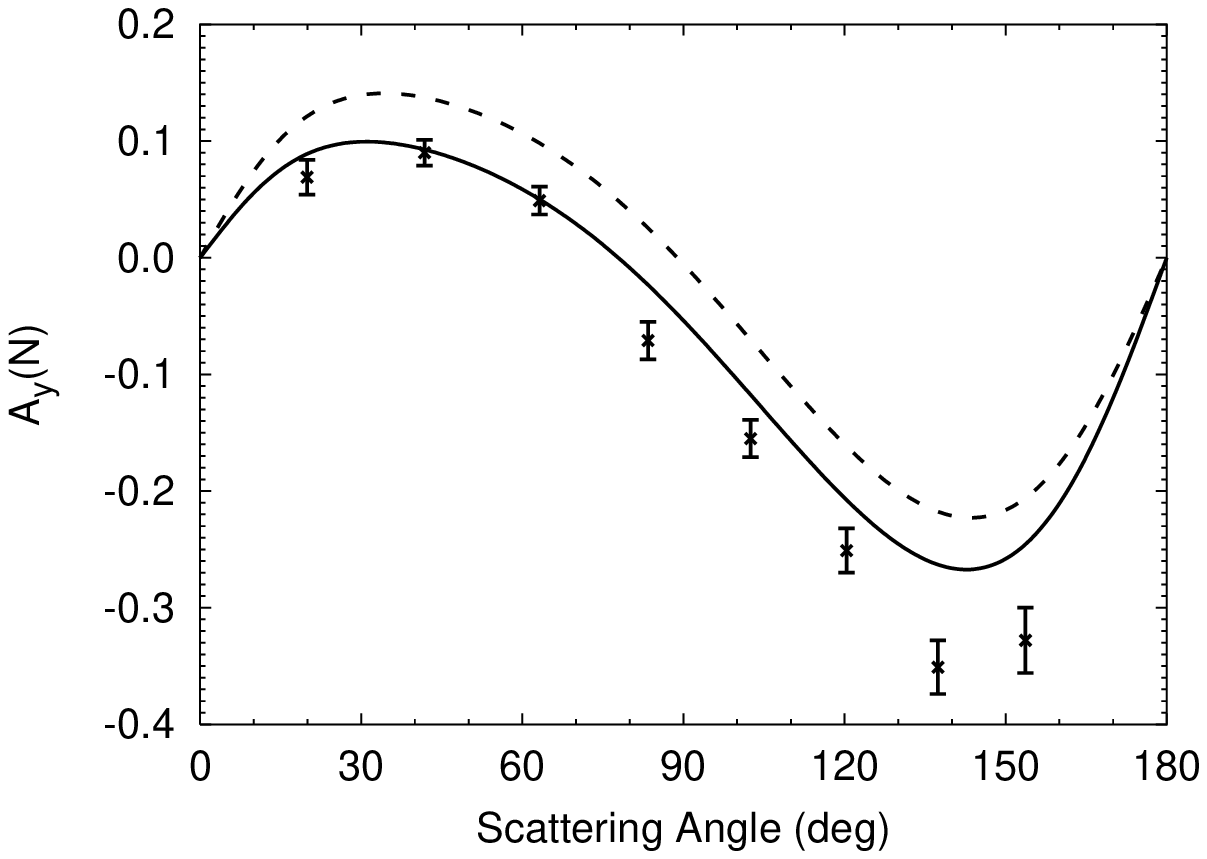} 
\caption{\label{fig:rc150}
Differential cross section and nucleon analyzing power of proton-deuteron 
radiative capture at 150~MeV  nucleon lab energy as function of the c.m. 
nucleon-photon scattering angle.
Results of the coupled-channel potential with $\Delta$-isobar excitation 
(solid curves) are compared with reference results of the purely nucleonic 
CD-Bonn potential (dashed curves).
The experimental data are from \Ref~\cite{pickar:87a}.}
\end{figure*}

Figures~\ref{fig:rc100} and \ref{fig:rc150}
present results for spin-averaged and spin-dependent observables of
radiative nucleon-deuteron capture at 100 and 150~MeV nucleon lab energy;
a result for 200~MeV will be shown later in \Sect~\ref{sec:rel}.
Results for the time-reversed two-body photo disintegration of the trinucleon
bound state are not shown separately. The energies are well above those
of \Ref~\cite{yuan:02a}, but remain below pion-production threshold.
 Control calculations at lower energies
indicate that the results of \Ref~\cite{yuan:02a} do not get any essential
physics change, though the hadronic interaction and the e.m. current are 
improved compared with \Ref~\cite{yuan:02a}.

There are noticeable $\Delta$-isobar effects on the considered observables,
especially on the nucleon analyzing power $A_y(N)$; 
it is described rather well with the inclusion of the $\Delta$ isobar.
Reference~\cite{yagita:03a} presents
 new  experimental data for the differential cross section
and deuteron vector analyzing power at 200~MeV deuteron lab energy,
corresponding to 100~MeV nucleon lab energy.
 There is a discrepancy between new~\cite{yagita:03a} and
old~\cite{pickar:87a} differential cross section data
in the maximum region; the new  data are in good agreement 
with our results including the $\Delta$ isobar. However, there is a clear 
disagreement between theoretical predictions and experimental data at
small scattering angles getting more pronounced at higher energies; 
one possible reason for that discrepancy is discussed in \Sect~\ref{sec:rel}.
There is also a modest beneficial $\Delta$-isobar effect on the deuteron
vector analyzing power $A_y(d)$.
The theoretical prediction for one deuteron tensor analyzing power, i.e., 
$A_{xx}$, is also given in \Fig~\ref{fig:rc100}; 
our motivation for showing $A_{xx}$ is the fact that an experiment
determining deuteron tensor analyzing powers is in progress~\cite{yagita:03a}.

Our results are qualitatively  consistent with those of 
\Refs~\cite{golak:00a,skibinski:03a,skibinski:03b}.

\subsection{\label{sec:dis3} Three-body photo disintegration 
of three-nucleon bound state }

\begin{figure}[!]
\includegraphics[scale=\scl]{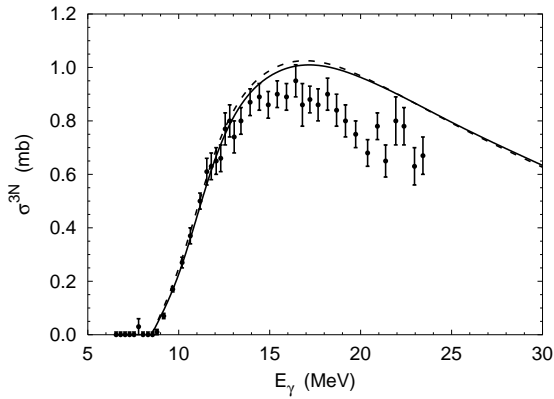} 
\caption{\label{fig:3Ntot}
Total ${}^3\mathrm{H}$ three-nucleon  photo disintegration cross section
as function of the photon lab energy $E_{\gamma}$. 
Results of the coupled-channel potential with $\Delta$-isobar excitation 
(solid curve) are compared with reference results of the purely nucleonic 
CD-Bonn potential (dashed curve).
The experimental data are from \Ref~\cite{faul:81a}.}
\end{figure}

\begin{figure}[!]
\includegraphics[scale=\scl]{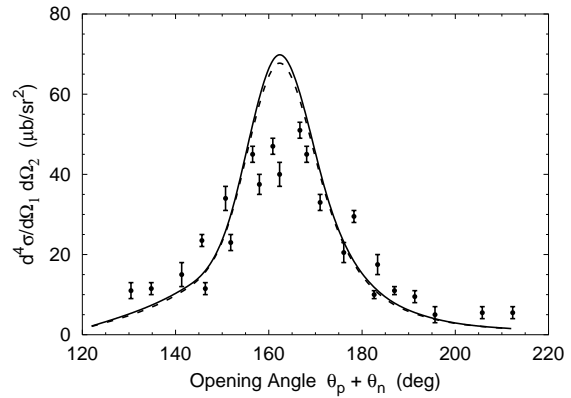} 
\caption{\label{fig:kolb}
The fourfold  differential cross section of the  ${}^3\mathrm{He}(\gamma,pn)p$ 
reaction at 85~MeV photon lab energy
as function of the $pn$ opening angle at $\theta_p = 81^{\circ}$.
Results of the coupled-channel potential with $\Delta$-isobar excitation 
(solid curve) are compared with reference results of the purely nucleonic 
CD-Bonn potential (dashed curve).
The experimental data are from \Ref~\cite{kolb:91a}.}
\end{figure}

\begin{figure}[!]
\includegraphics[scale=\scl]{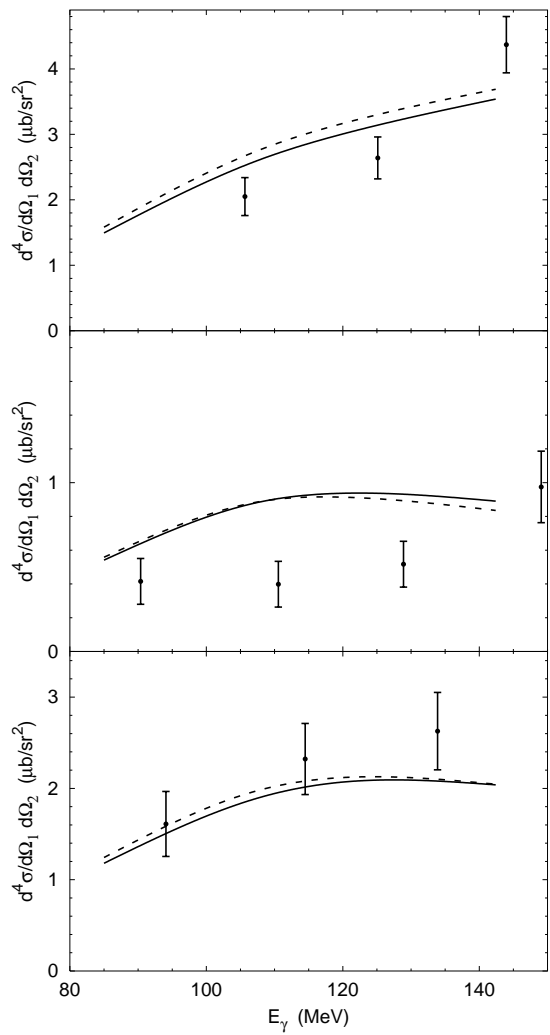} 
\caption{\label{fig:sarty}
The fourfold  differential cross section of the  ${}^3\mathrm{He}(\gamma,pp)n$ 
reaction as function of the photon lab energy $E_{\gamma}$ in various 
kinematical configurations:
$(81.0^{\circ},81.3^{\circ},180.0^{\circ})$ (top), 
$(92.2^{\circ},91.4^{\circ},180.0^{\circ})$ (middle),
and average of $(81.5^{\circ},90.8^{\circ},180.0^{\circ})$ and
$(91.7^{\circ},80.9^{\circ},180.0^{\circ})$ (bottom).
Results of the coupled-channel potential with $\Delta$-isobar excitation 
(solid curve) are compared with reference results of the purely nucleonic 
CD-Bonn potential (dashed curve).
The experimental data are from \Ref~\cite{sarty:93a}.}
\end{figure}

Experimental data for three-nucleon breakup are much scarcer than for 
two-body photo disintegration. To the best of our knowledge, there are no fully
exclusive experimental data in the considered energy regime; 
we therefore show in \Figs~\ref{fig:3Ntot} -- \ref{fig:sarty}
our predictions  for inclusive and semi-exclusive observables
and compare them with existing experimental data.
Figure~\ref{fig:3Ntot} shows our results for  the total ${}^3\mathrm{H}$ 
three-nucleon photo disintegration cross section in the low energy region;
there is no significant $\Delta$-isobar effect. 
In contrast, \Ref~\cite{skibinski:03b} sees a larger three-nucleon force effect
for this observable; this discrepancy is partly due to a larger three-nucleon
force effect on trinucleon binding and subsequent scaling and partly due to
a different computational strategy as discussed in \Sect~\ref{sec:Kin}.
Figures~\ref{fig:kolb} -- \ref{fig:sarty} show semi-exclusive fourfold 
differential cross sections of ${}^3\mathrm{He}$ photo disintegration
at higher energies; they are obtained from
the fivefold differential cross section~\eqref{eq:d5s-av} by integrating 
over the kinematical curve $S$. Again, the $\Delta$-isobar effect for those
particular observables appears rather small, smaller than the experimental error
bars. There is also disagreement between theoretical predictions and experimental
data in some kinematical regimes which  in part may be caused by experimental
conditions, e.g., by finite geometry, not taken into account in our calculations.

\begin{figure}[!]
\includegraphics[scale=\scl]{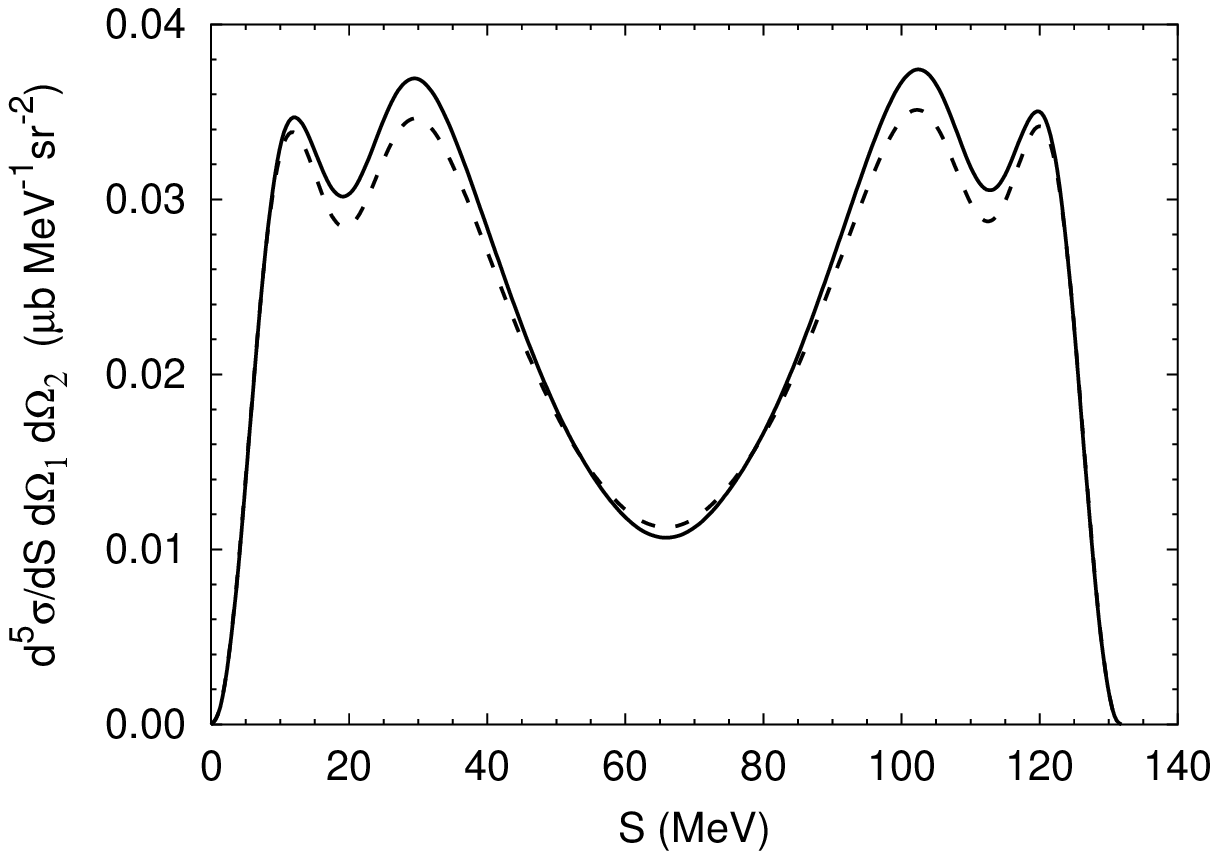} \\
\includegraphics[scale=\scl]{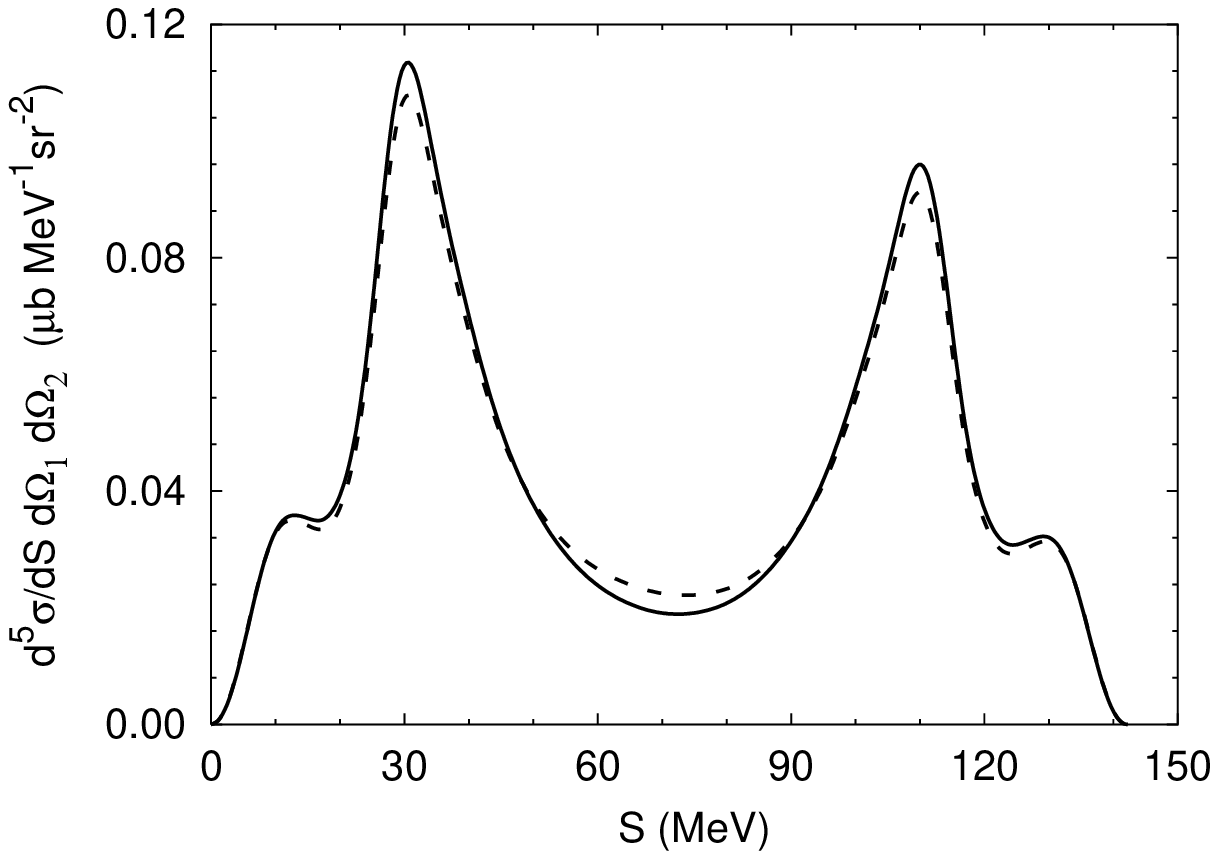} 
\caption{\label{fig:d5s}
The fivefold differential cross section  of three-nucleon photo disintegration
at 120~MeV  photon lab energy as function of the arclength $S$ along 
the kinematical curve  for  configuration 
$(92.2^{\circ},91.4^{\circ},180.0^{\circ})$ on the top
and  $(81.5^{\circ},90.8^{\circ},180.0^{\circ})$ on the bottom.
Results of the coupled-channel potential with $\Delta$-isobar excitation 
(solid curve) are compared with reference results of the purely nucleonic 
CD-Bonn potential (dashed curve).}
\end{figure}

Finally, in \Fig~\ref{fig:d5s} we show fully exclusive sample
 fivefold differential cross sections of three-nucleon
photo disintegration at 120~MeV photon lab energy for two kinematical 
configurations which were shown semi-exclusively in \Fig~\ref{fig:sarty};
even at that higher energy the $\Delta$-isobar effect is rather mild.

\section{\label{sec:shortcomings} Shortcomings of the description}

The present description of photo reactions is with respect to the dynamic 
input, i.e., with respect to the hadronic interaction and to the e.m. current,
 and with respect to the scope of applications a substantial improvement 
compared with \Ref~\cite{yuan:02a}. But it is still not a unique and in itself
consistent description.
We are unable to repair the existing deficiencies. However, this section 
points those shortcomings out and tries at least to estimate their size.
We identify three different problem areas.

\subsection{\label{sec:NRKin} Shortcomings of the theoretical form of the
cross section}

Our standard strategy uses the nonrelativistic form~\eqref{eq:d5Sn}
for cross sections; this 
choice appears to be consistent with the underlying two-baryon dynamics,
though inconsistent with the experimental relativistic kinematics.
We therefore compare results obtained from \Eqs~\eqref{eq:d5Sn} with
corresponding ones obtained from the relativistic form of the cross section 
\eqref{eq:d5Sr}  which uses relativistic kinetic 
energies  for the Lorentz-invariant phase space element
\eqref{eq:dLips}  and the kinematic locus~\eqref{eq:Scurve}
combined with the nonrelativistic matrix element \eqref{eq:Mampl}.
The comparison is possible for observables in fully exclusive reactions.

\begin{figure}[!]
\includegraphics[scale=\scl]{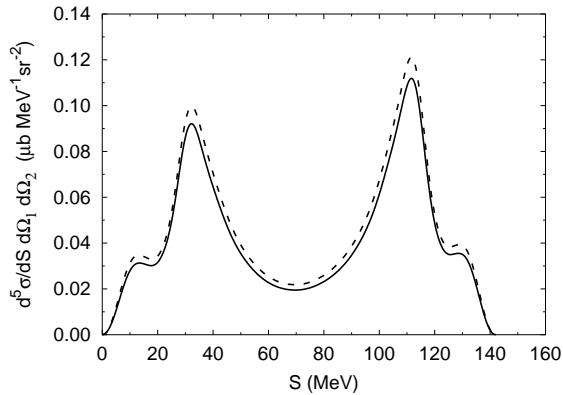} 
\caption{\label{fig:kin3}
Differential cross section  of three-nucleon photo disintegration
at 120~MeV  photon lab energy as function of the arclength $S$ along 
the kinematical curve  for  configuration 
$(91.7^{\circ},80.9^{\circ},180.0^{\circ})$.
Results of the coupled-channel potential with $\Delta$-isobar excitation
based on nonrelativistic phase space (solid curve) 
according to \Eqs~\eqref{eq:d5Sn} are compared 
with results based on relativistic phase space (dashed curve)
according to \Eqs~\eqref{eq:d5Sr}.}
\end{figure}

The difference between those aspects of relativistic and nonrelativistic
kinematics is minor for all considered observables of radiative capture,
i.e., less than 1\%, but more significant, i.e., up to  10\%,  for
three-nucleon photo disintegration as shown in \Fig~\ref{fig:kin3}.
In all considered cases, the relativistic and nonrelativistic kinematical curves
\eqref{eq:Scurve} are very close to each other, e.g., for the configuration
of \Fig~\ref{fig:kin3} the distance between them in the $E_1-E_2$ plane is
0.5~MeV at most, and their total arclengths are 140.2~MeV and 142.0~MeV,
respectively. For the comparison the nonrelativistic results in 
\Fig~\ref{fig:kin3} are scaled down to the relativistic arclength
by the factor 140.2/142.0; the shown change, however, is due to the 
difference in the phase space factors $\mathrm{fps}$ of 
 \Eqs~\eqref{eq:fps} and \eqref{eq:fpsn}.

We emphasize: The  effect indicated in this subsection does not represent the 
true difference between
nonrelativistic quantum mechanical and fully relativistic quantum field 
theoretical results, but it may indicate the order of magnitude of the
shortcomings of nonrelativistic calculations.
In the light of the accuracy of present day data, this shortcoming
of the theoretical description is rather inconsequential.

\subsection{Shortcomings of the dynamics}

\subsubsection{\label{sec:Kin} Nonunique choice of kinematics}

\begin{figure}[!]
\includegraphics[scale=\scl]{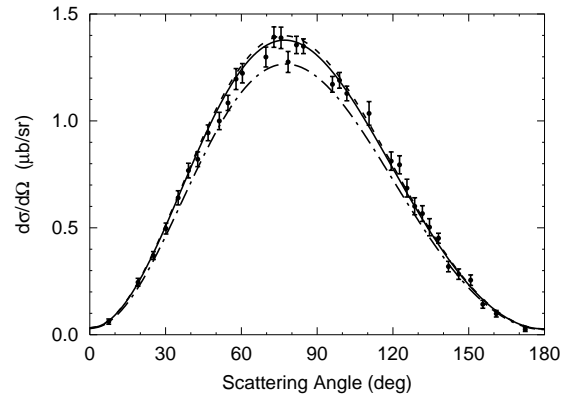} 
\caption{\label{fig:offsh}
Differential cross section  of proton-deuteron radiative capture at 
19.8~MeV deuteron lab 
energy as function of the c.m. nucleon-photon scattering angle.
Results of the coupled-channel potential with $\Delta$-isobar excitation 
derived from the standard approach (solid curve) are compared with results 
of option three which uses experimental energies for both initial and 
final states, but the matrix element~\eqref{eq:MamplN} is off-shell
(dashed-dotted curve). The results of option two are rather close to the
solid curve. In order to appreciate the effect of the nonunique choice of 
kinematics in relation to the size of the $\Delta$-isobar effect, 
results of a standard calculation with the purely nucleonic reference potential
 are also given as dashed curve.
The experimental data are from \Ref~\cite{belt:70a}.}
\end{figure}

Our computational strategy in choosing the kinematics for the matrix element
$\langle s_f | M | s_i \rangle$ is described in \Sect~\ref{sec:smat}. 
$\langle s_f | M | s_i \rangle$ is calculated in the c.m. system.
We opt to let the experimental beam energy determine the energy of hadronic
nucleon-deuteron state  in radiative capture and the energy of the hadronic 
two-body and three-body final states in photo disintegration exactly. Since the 
trinucleon model binding energy is not the experimental one and the
kinematics  is nonrelativistic for baryons when calculating 
$\langle s_f | M | s_i \rangle$, the energy of the  photon does not have 
the experimental value when assuming energy conservation.
At very low energies the deviation can get as large as 10\%, whereas at 
higher  energies considered in this paper it remains around 1 - 2\%.
In contrast, in a second option
we could  let the experimental beam energy determine the
c.m. photon energy exactly; then the energies of the hadronic nucleon-deuteron
and three-nucleon states are not experimental ones. A third option may use 
experimental energies for both initial and final states, but then the matrix element
determining physical amplitudes is slightly off-shell; this is the computational
strategy of \Refs~\cite{skibinski:03a,skibinski:03b}. The difference
in results between those three choices is minor at higher energies, i.e.,
above 100~MeV nucleon lab energy,
for all considered observables in all considered kinematical regimes.
However, there are differences up to 10\% for observables at low energies. 
There, the observed
$\Delta$-isobar effect depends strongly on the choice of computational strategy.
An example is shown in \Fig~\ref{fig:offsh}.

\subsubsection{\label{sec:CoulB} Omission of Coulomb interaction between protons}

\begin{figure}[!]
\includegraphics[scale=\scl]{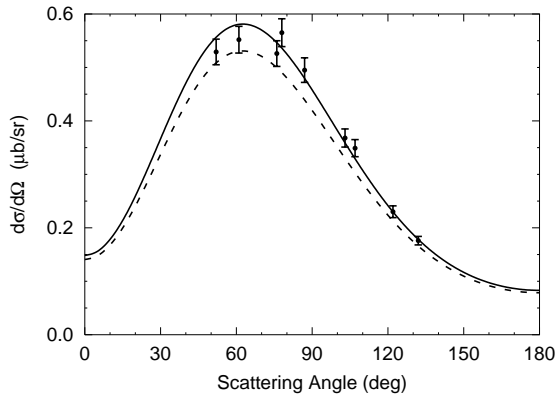} 
\caption{\label{fig:coul}
Differential cross section  of proton-deuteron radiative capture at 
95~MeV deuteron lab 
energy as function of the c.m. nucleon-photon scattering angle.
Results of the coupled-channel potential with $\Delta$-isobar excitation 
derived from the standard approach
(solid curve) are compared with results including the Coulomb interaction in 
the three-nucleon bound state (dashed curve).
The experimental data are from \Ref~\cite{pitts:88a}.}
\end{figure}

We are unable to include the Coulomb interaction in the three-nucleon 
scattering states. In contrast, the selected inclusion of the Coulomb interaction
in the trinucleon bound state is easily possible, but this inclusion
creates an additional inconsistency: Initial
and final hadronic states become eigenstates of different Hamiltonians,
and, strictly speaking, the Siegert form of the current operator
 is not applicable. Nevertheless, we do such an inconsistent calculation
which \Refs~\cite{skibinski:03a,skibinski:03b} chooses to do as
standard calculation, 
in order to estimate the effect of the omitted Coulomb interaction
at least partially. The inclusion of the Coulomb interaction
in the trinucleon bound state systematically reduces the spin-averaged cross 
sections; in contrast, spin observables appear to be almost unaffected.
A characteristic result is shown in \Fig~\ref{fig:coul}. Even at higher
energies the observed Coulomb effect may be of the same order of magnitude as
the full $\Delta$-isobar effect; however, it is not clear
whether the indicated effect represents a true Coulomb effect or just the 
inconsistency between bound and scattering states.

\subsection{Shortcomings of the e.m. current}
\subsubsection{\label{sec:Smec} Lack of current conservation}

\begin{figure}[!]
\includegraphics[scale=\scl]{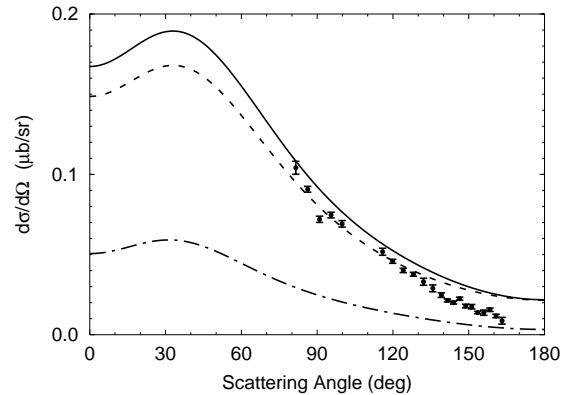} 
\caption{\label{fig:s-mec}
Differential cross section  of proton-deuteron 
radiative capture at 190~MeV  nucleon lab energy as function of the c.m. 
nucleon-photon scattering angle.
Results of the coupled-channel potential with $\Delta$-isobar excitation 
derived from the Siegert approach for electric multipoles
(solid curve) are compared with results based on the explicit use
of meson-exchange currents (dashed curve). In order to appreciate the size
of the two-baryon current contribution, the results of a non-Siegert calculation
with one-baryon currents only are also given as dashed-dotted curve.
The experimental data are from \Ref~\cite{messchendorp:00a}.}
\end{figure}

The potentials CD Bonn and CD Bonn + $\Delta$ used in this paper
have nonlocal structures, whereas the e.m. current, given explicitly
in Appendix~\ref{app:current} is employed in a local nonrelativistic 
form. Thus, the continuity equation is not fulfilled for the current. 
As measure for this deficiency predictions are compared based on two
different approaches for the electric multipoles, i.e., (1)
the standard calculation with the Siegert operator accounting for the 
two-baryon currents implicitly by assumed current conservation and (2) the 
explicit use of the meson-exchange currents for all of the electric multipoles.
The discrepancy between those two calculations measures the importance of the
existing lack of current  conservation; indeed the violation can be 
significant as \Fig~\ref{fig:s-mec} proves. We believe that calculations
with the Siegert form of the current operator repair the violation of current
conservation in part; we therefore employ  the Siegert form of the current
operator in our standard calculational strategy. 

However, at this stage it is useful to discuss the lack of current 
conservation in more detail:

1) 
The $\sigma$, $\rho$ and $\omega$ exchanges yield a spin-orbit interaction. 
That spin-orbit interaction makes, even in local approximation and even for 
isoscalar-meson exchanges, a contribution to the continuity equation. 
 The corresponding spin-orbit
contribution to the exchange current is derived in local form~\cite{riska:85a}.
There are also additional contributions to the  $\rho$-exchange current
\cite{adam:89a} not taken into account in our standard calculations.
All that contributions are implicitly contained in the Siegert-part of 
the electric multipoles. In the tentative calculations described in this 
paragraph they are
 used explicitly for the non-Siegert part of the electric  multipoles
and for the magnetic multipoles. There, that sample  contributions
yield only small corrections for observables considered in this paper,
seen mostly in spin observables; however, even for them the corrections
are of the order of 2\% at most. We therefore conclude that
spin-orbit contributions and additional $\rho$-exchange currents
 can quite safely be neglected in the Siegert form 
of the current operator when calculating the photo reactions of this paper.
However, they are more important for calculations based fully on explicit 
exchange currents and they therefore make a non-negligible contribution 
to the difference seen in  \Fig~\ref{fig:s-mec}.

2)
The employed potentials have explicit nonlocal structures. That explicit 
nonlocality makes a contribution to the two-baryon exchange current.

We compared the results of the following models for the purely nucleonic 
potential. The models are based on $\pi$, $\rho$ and $\omega$ exchanges
and on a scalar isoscalar and a scalar isovector $\sigma$ exchange;
one model uses the nonlocal structures of CD-Bonn type and the other their
local approximations. Both models are tuned to deuteron binding and to
${}^1S_0$ and ${}^3S_1 - {}^3D_1$ phase shifts only. Though nonrealistic
models, both usually predict observables in qualitative agreement with the
realistic descriptions. When employing the local  potential model we obtain
identical results for calculations based on the Siegert form of the current
operator and calculations based on explicit exchange-current contributions
to all multipoles. However, when employing the nonlocal potential model,
the results can differ substantially; the difference can be as large as seen
in  \Fig~\ref{fig:s-mec} for the coupled-channel potential CD Bonn + $\Delta$.
We conclude, the explicit nonlocality of the employed potentials is a 
significant source for current non-conservation. Future calculations should
attempt to design nonlocal exchange-current contributions consistent with
the nonlocality of the underlying baryon-baryon potentials.

3) 
The employed potentials have an implicit nonlocality due to the general
partial-wave dependence  of the meson exchanges. That partial-wave dependence 
is slight for the $\pi$, $\rho$ and $\omega$ exchanges, but substantial for
the $\sigma$ exchange. That  implicit nonlocality makes a contribution to 
the two-baryon exchange current.

The nucleonic CD Bonn as well as CD Bonn + $\Delta$ show a small partial-wave
dependence in $\pi$ and in $\rho$ exchange. Fitting CD Bonn with partial wave
independent $\pi$ and  $\rho$ exchanges decreases the quality of the fit to data
only by very little; $\chi^2/\mathrm{datum}$ increases from 1.02 to 1.03. 
When comparing observables of the hadronic processes and of the photo reactions
of this paper for both potentials, no distinguishable difference is found in
plots. We conclude: The implicit nonlocality arising from the partial-wave
dependence in $\pi$ and  $\rho$ exchange of CD Bonn and CD Bonn + $\Delta$
is of no consequence for the prediction of observables.

The local model used for the discussion of problem 2) is modified to simulate
the partial-wave dependence of $\omega$ exchange in the nucleonic CD Bonn,
it is retuned as under 2). The $\omega$ exchange  is taken to be without
hadronic cutoff in the ${}^1P_1$ partial wave as in CD Bonn; 
this partial-wave dependence violates  current conservation. 
However, the observed difference between
calculations based on the Siegert form of the current
operator and calculations based on explicit exchange-current contributions
to all multipoles is much smaller than that shown in \Fig~\ref{fig:s-mec}.
We conclude: The implicit nonlocality arising from the partial-wave dependence
in the  $\omega$-exchange of CD Bonn and CD Bonn + $\Delta$ is of no real 
consequence for the prediction of observables.

With respect to the partial-wave dependence of $\sigma$ exchange the local
model used for the discussion of problem 2) is studied. We concentrate on
the difference of $\sigma$ exchange between isospin singlet and triplet
partial waves, i.e., on the effective isovector nature of the $\sigma$ meson
introduced in the model and in CD Bonn and CD Bonn + $\Delta$.
Furthermore, even if the  $\sigma$ exchange  were truly an isoscalar one
in purely nucleonic potential, the explicit treatment of the $\Delta$ isobar
in the coupled-channel extension introduces an isovector correction:
The employed coupled-channel potential CD Bonn + $\Delta$, acting in 
isospin-triplet partial waves, has a weakened $\sigma$ exchange compared to
the purely nucleonic CD Bonn; part of the intermediate range attraction 
simulated by $\sigma$ exchange is taken over by $\Delta$-isobar excitation
in the coupled-channel approach. Thus, the $\Delta$-isobar current has to be 
supplemented by changed $\sigma$-exchange current. Omitting the 
$\sigma$-meson contribution to the exchange current, quite significant 
differences, comparable to that of \Fig~\ref{fig:s-mec}, arise for observables
of the photo reactions in this paper between calculations based on the 
Siegert form of the current and on the full explicit  exchange-current
 contributions. In contrast, the explicit $\sigma$-meson exchange-current
contributions to the non-Siegert part of the electric multipoles and to the
magnetic multipoles remain small. We arrive to qualitatively the same results
when including the $\sigma$-meson exchange-current for CD Bonn and 
CD Bonn + $\Delta$ with the $\sigma$-meson parameters of $S$-waves.
We conclude: Though the partial-wave dependence of the $\sigma$-meson exchange
is a significant source of current nonconservation, the standard calculation
based on the Siegert form of the current for part of the electric multipoles
 and on explicit exchange-current contributions to all other multipoles
appears to be quite a reliable calculational scheme.

4)
The employed potentials are charge dependent. The charge dependence of the 
interaction is due to the charge dependence of the parameters of exchanged
 $\pi$, $\rho$ and $\sigma$ mesons and due to 
the charge dependence of the  nucleonic masses. The isospin structure
of the charge-dependent potential contributions is given in terms of the
baryonic isospin projections; thus, that isospin dependence, giving
rise to  charge dependence, does not require an exchange current by itself;
it only does so, if its potential forms were nonlocal. In the case of the 
employed potentials it is so indeed, but that explicit nonlocality was already
discussed in problem 2). The diagonal  $\pi$- and $\rho$-exchange
contributions to the exchange current should be  built from the meson
parameters of the charged mesons. The nondiagonal  $\pi$ and $\rho$ exchanges
are carried  by the mesons of all charges. However, our standard
calculation uses averaged meson parameters and an averaged 
nucleon masses for all meson-exchange currents; it was checked that both 
calculational simplifications are without any consequence for the observables of 
this paper.

From this lengthy, but we think necessary discussion of the problems 1) to 4)
we conclude for the calculations of this paper: When the Siegert form of the 
current is used for part of the electric multipoles and 
explicit exchange-current contributions to all other
multipoles in the operator form of Appendix~\ref{app:current}, 
the implicit nonlocality of CD Bonn and CD Bonn + $\Delta$
arising from the partial-wave dependence  of the meson exchanges is
without consequences for prediction. In contrast, the explicit nonlocality
of CD Bonn and CD Bonn + $\Delta$, also responsible for current nonconservation,
is of serious concern; its consequence on the non-Siegert parts of the current
could not be estimated yet by any of our models. Still, we believe that our
standard calculation, based on the Siegert form of the current,
 effectively corrects the current nonconservation and is
therefore quite reliable for the observables of photo reactions considered
in this paper.

\subsubsection{\label{sec:frmdep} Lack of covariance}

If a fully covariant description of dynamics were available,
the current matrix element  $ \langle s_f | M | s_i \rangle$ 
 were a Lorentz scalar and could therefore be calculated in any 
frame with identical results. However, our description of hadron dynamics is
nonrelativistic, and the results therefore  are frame-dependent. We investigate
that frame dependence calculating the same matrix elements in lab and in 
c.m. frames, i.e., in the rest frames of the initial and final three-nucleon 
systems. The two frames differ by  the three-nucleon 
total momentum and by the photon momentum. 
However, we found that for the observables considered in this paper
the frame dependence is minor and at present of no real theoretical concern;
we do not document that finding, since the differences 
are only hardly  seen in plots.

\subsubsection{\label{sec:rel} Higher order contributions 
to the current operator in $(k/m_N)$ expansion}

\begin{figure}[!]
\includegraphics[scale=\scl]{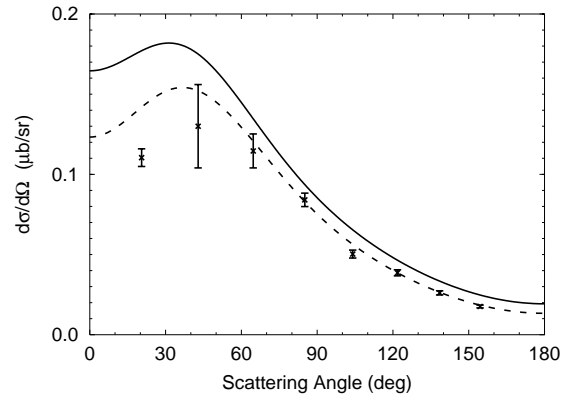} 
\caption{\label{fig:relcorr}
Differential cross section  of proton-deuteron 
radiative capture at 200~MeV  nucleon lab energy as function of the c.m. 
nucleon-photon scattering angle.
Results of the coupled-channel potential with $\Delta$-isobar excitation 
derived from our standard approach
(solid curve) are compared with results including relativistic one-nucleon
charge corrections (dashed curve).
The experimental data are from \Ref~\cite{pickar:87a}.}
\end{figure}

In the standard calculational scheme the Siegert form of the current operator
is used together with explicit meson-exchange contributions not accounted for
by the Siegert part. The charge density operator in the Siegert part is of 
one-baryon nature and is taken to be nonrelativistic in the standard calculations.
However, the one-baryon purely nucleonic charge density  
operator has relativistic corrections of order $(k/m_N)^2$.
Contributions to   the nucleon-$\Delta$
transition charge density and  to the two-nucleon charge density,
used in \Ref~\cite{henning:92a} for calculation of trinucleon elastic
charge form factors, are also included; both are of the relativistic order 
$(k/m_N)^2$.
The resulting special relativistic corrections, taken into account in this paper,
  reduce the cross sections; they appear beneficial; 
a characteristic result is shown in \Fig~\ref{fig:relcorr}.
The effect shown there is dominated by the one-nucleon charge-density 
correction;  the two-nucleon charge-density  shows noticeable 
effects in  some spin observables, whereas the  nucleon-$\Delta$
transition charge appears to be insignificant for all calculated
observables of this paper.
Correspondly large corrections of the same origin were also found in 
photo reactions on the deuteron~\cite{arenhoevel:99a}.
Thus, the results of this subsection are not surprising.
The current corrections  of this subsection should be included in future 
calculations of e.m. reactions.

\section{\label{sec:concl} Summary and conclusions}

The paper improves our preliminary description of photo reactions in the
three-nucleon system~\cite{yuan:02a}; 
the present description includes three-body photo
disintegration. The hadronic interaction is based on CD Bonn and its 
realistic coupled-channel extension CD Bonn + $\Delta$.
The initial and final hadronic states are calculated without separable
expansion of the underlying interaction. The contributions to the e.m. current
correspond to the hadronic interaction, though full current conservation
 could not be achieved.

The paper isolates the $\Delta$-isobar effects on the considered observables.
Besides $\Delta$-isobar effects of effective two-nucleon nature the 
$\Delta$ isobar yields an effective three-nucleon force of the Fujita-Miyazawa
type and of the Illinois pion-ring type; meson exchanges other than pion
exchange are included. The exchange currents mediated by the $\Delta$ isobar 
are of effective two-nucleon and three-nucleon nature; they are structurally
consistent with corresponding hadronic contributions; they are predominantly
due to the transition contributions of \Fig~\ref{fig:Jnd} and due to the one-baryon
part of \Fig~\ref{fig:Jdd}; the diagonal two-baryon contributions  
of \Fig~\ref{fig:Jdd} connect small wave function components and are found 
to be quantitatively entirely irrelevant.  In the considered
observables the hadronic and the e.m. $\Delta$-isobar effects are
intertwined; they are not separated; their total effects are not very important
given the scarcity of data, often still carrying large error bars.
The $\Delta$-isobar effects are more pronounced at higher energies;
they are somehow smaller than the irreducible three-nucleon force effects
 of \Refs~\cite{skibinski:03a,skibinski:03b}; nevertheless,
qualitatively both effects are  quite similar.
In contrast to the three-nucleon force effects of 
\Refs~\cite{skibinski:03a,skibinski:03b} we see very small 
$\Delta$-isobar effects at low energies; the reason, at least in part, is due
to the different choice of kinematics for correcting the theoretical failure
in accounting for three-nucleon binding. 

\begin{acknowledgments}
The authors thank J.~Golak, J.~Messchendorp and R.~Skibi\'nski for providing 
them with the experimental data.
A.D. acknowledges a valuable DAAD grant for graduate studies at the
University of Hannover.
L.P.Y. and P.U.S. are supported by the DFG grant Sa 247/25,
J.A. by the grant GA CzR 202/03/0210 and A.C.F. by the grant
POCTI/FNU/37280/2001. 
The numerical calculations were performed at Regionales Rechenzentrum 
f\"ur Niedersachsen.
\end{acknowledgments}


\begin{appendix}
\section{\label{app:current} Coupled-channel current operators}

Equation~\eqref{eq:JpqK} defines the general momentum space form of the e.m.
current $J^{\mu}(\vec{Q})$ in the Jacobi coordinates of the three-particle basis.
In contrast, this appendix gives its employed one-baryon and two-baryon parts,
i.e., $J^{\mu}(\vec{Q}) = J^{[1]\mu}(\vec{Q}) + J^{[2]\mu}(\vec{Q}) $, in
respective one-particle and two-particle bases. 
We keep the three-momentum transfer $\vec{Q}$ and not the four-momentum transfer 
$Q$ as independent variable since usually $Q_0$ is determined by the 
three-momenta of the involved baryons. Despite that strategy, 
$Q^2 = \vec{Q}^2 - Q_0^2$ is taken to be zero in all e.m. form factors for
photo reactions. The step from the single-particle 
representation of the current contributions to the three-particle Jacobi momenta
is straightforward~\cite{oelsner:phd} and not repeated here. The objective
of this appendix is the definition of the used input for the current.

\subsection{One-baryon operators in nonrelativistic order}

The momentum-space matrix elements of the one-baryon current operator have 
the general form
\begin{gather}
  \langle \vec{k}' B' | J^{[1]\mu}(\vec{Q}) | \vec{k} B \rangle =
  \delta( \vec{k}' - \vec{Q} - \vec{k}) 
  j^{[1]\mu}_{B'B}(\vec{Q},\vec{k}',\vec{k})
\end{gather}
with $\vec{k}' \; (\vec{k})$ and $\vec{Q}$ being the final (initial) single-baryon 
momentum and the four-momentum transfer by the photon, respectively, and
 $B' \; (B)$  being $N$ or $\Delta$ depending on the baryonic content of the 
final (initial) state. All components of 
$j^{[1]\mu}_{B'B} (\vec{Q},\vec{k}',\vec{k})$ are still operators in spin and 
isospin space; the spin (isospin) operators of the nucleon, $\Delta$ isobar and the
nucleon-$\Delta$ transition are denoted by
$\vecg{\sigma}$ $(\vecg{\tau})$, $\vecg{\sigma}_{\Delta}$ $(\vecg{\tau}_{\Delta})$
and $\vec{S}$ $(\vec{T})$, respectively. 
The one-baryon charge density and spatial current operators, diagrammatically
defined in \Figs~\ref{fig:Jnn} - \ref{fig:Jdd} and used in the calculations
of this paper, are listed below:

\begin{subequations} \label{eq:J1b}
  \begin{align} \label{eq:J1n0}
    \rho^{[1]}_{NN} (\vec{Q},\vec{k}',\vec{k}) = {}& e(Q^2), \\
    \vec{j}^{[1]}_{NN} (\vec{Q},\vec{k}',\vec{k}) = {}& \frac{1}{2m_N} 
    \big \{ e(Q^2) [\vec{k}'+\vec{k}] 
    \nonumber \\ & 
    + [e(Q^2)+\kappa(Q^2)] [i\vecg{\sigma} \times \vec{Q}] \big \}, \\
    \vec{j}^{[1]}_{\Delta N} (\vec{Q},\vec{k}',\vec{k}) = {}& \frac{1}{2m_{\Delta}} 
    g^{\mathrm{M1}}_{\Delta N}(Q^2) [i\vec{S} \times \vec{Q}] T_z, \\
    \label{eq:J1d0}
    \rho^{[1]}_{\Delta \Delta} (\vec{Q},\vec{k}',\vec{k}) = {}& 
    g^{\mathrm{E0}}_{\Delta}(Q^2) , \\
    \vec{j}^{[1]}_{\Delta \Delta} (\vec{Q},\vec{k}',\vec{k}) = {}& 
    \frac{1}{2m_{\Delta}}
    \big \{ g^{\mathrm{E0}}_{\Delta}(Q^2) [\vec{k}' + \vec{k}]
    \nonumber \\ &
    +  g^{\mathrm{M1}}_{\Delta}(Q^2) [i \vecg{\sigma}_{\Delta} \times \vec{Q}]
    \big \}.
\end{align}
\end{subequations}
The nucleonic e.m. form factors of \Eqs~\eqref{eq:J1b} are parametrized as
 linear combinations of the isospin-dependent Dirac and Pauli form factors
$f_1(Q^2)$ and $f_2(Q^2)$, which at zero four-momentum transfer $Q^2$ are
the charge and the anomalous magnetic moment of the nucleon, i.e.,
\begin{subequations} \label{eq:ek}
  \begin{align} \label{eq:e}
    e(Q^2) = {}& \frac12 \big[ f_1^{\mathrm{IS}}(Q^2) +  
      f_1^{\mathrm{IV}}(Q^2) \tau_z \big], \\
    \kappa(Q^2) = {} & \frac12 \big[ f_2^{\mathrm{IS}}(Q^2) +  
      f_2^{\mathrm{IV}}(Q^2) \tau_z \big],
  \end{align}
\end{subequations}
the superscripts IS and IV  denote their isoscalar and isovector parts, 
respectively.
 The e.m. form factors related to the $\Delta$-isobar are parameterized 
\cite{hajduk:83a,carlson:86a,lin:91a} by

\begin{subequations} \label{eq:dff}
  \begin{align} \label{eq:dffa}
    g^{\mathrm{M1}}_{\Delta N}(Q^2) = {}& \frac{m_{\Delta}}{m_N}
    \frac{\mu_{\Delta N}}{(1+Q^2/\Lambda^2_{\Delta N,1})^2
      (1+Q^2/\Lambda^2_{\Delta N,2})^{1/2}}, \\
    g^{\mathrm{E0}}_{\Delta}(Q^2) = {}& \frac12 
    \big \{ f_1^{\mathrm{IS}}(Q^2) +  f_1^{\mathrm{IV}}(Q^2) - Q^2/(4m_N^2) 
    \nonumber \\ & \times
      [f_2^{\mathrm{IS}}(Q^2) +  f_2^{\mathrm{IV}}(Q^2)] \big \}
      \, \frac12 (1 + \vecg{\tau}_{\Delta z}), \\
    g^{\mathrm{M1}}_{\Delta}(Q^2) = {}& \frac{m_{\Delta}}{6 m_N}
    \frac{\mu_{\Delta}}{(1+Q^2/\Lambda^2_{\Delta})^2} \,
    \frac12 (1 + \vecg{\tau}_{\Delta z}).
  \end{align}
\end{subequations}
The values of the parameters are $\mu_{\Delta N} = 3\,\mu_N$, 
 $\mu_{\Delta} = 4.35\,\mu_N$, $\mu_N$ being the nuclear magneton,
$\Lambda_{\Delta N,1} = \Lambda_{\Delta} = 0.84\;\mathrm{GeV}$ and
$\Lambda_{\Delta N,2} = 1.2\;\mathrm{GeV}$.

\subsection{Two-baryon operators in nonrelativistic order}
The matrix elements of the two-baryon current operator have the general form
\begin{gather}
  \begin{split} 
    \langle \vec{k}_1' \vec{k}_2' B' & |J^{[2]\mu}_{\alpha}(\vec{Q})| 
    \vec{k}_1 \vec{k}_2 B \rangle 
     \\ = {} & 
    \delta( \vec{k}_1' + \vec{k}_2' - \vec{Q} - \vec{k}_1 - \vec{k}_2) 
     \\ & \times
    j^{[2]\mu}_{\alpha B'B}(\vec{Q},\vec{k}_1'-\vec{k}_1, \vec{k}_2'-\vec{k}_2)
  \end{split} 
\end{gather}
with $\vec{k}_i' \; (\vec{k}_i)$  being the final (initial) single-baryon momenta; 
$\alpha$ denotes the exchanged meson or the two mesons in case of nondiagonal 
currents); the baryonic contents $B'$ and $B$ being  $N \; (\Delta) $ 
correspond to the two-nucleon (nucleon-$\Delta$-isobar) states. All components 
$ j^{[2]\mu}_{\alpha B'B}(\vec{Q},\vec{k}_1'-\vec{k}_1, \vec{k}_2'-\vec{k}_2)$
are still operators in spin and isospin space.
The two-baryon spatial current operators, diagrammatically
defined in \Figs~\ref{fig:Jnn} - \ref{fig:Jdd} and used in the calculations
of this paper, are listed below:
\begin{widetext}
\begin{subequations} \label{eq:J2n}
  \begin{align} 
    \vec{j}^{[2]}_{\pi NN}(\vec{Q},\vec{p}_1, \vec{p}_2) = {}&
    -  f_1^{\mathrm{IV}}(Q^2) \big\{ [i \vecg{\tau}_1 \times \vecg{\tau}_2]_z
    F^{\mathrm{con}}_{\pi NN} (\vec{p}_2^2)(\vec{p}_2 \cdot \vecg{\sigma}_2)
    \vecg{\sigma}_1 + (1 \leftrightarrow 2) \big \} \nonumber \\
    {}& + f_1^{\mathrm{IV}}(Q^2) [i \vecg{\tau}_1 \times \vecg{\tau}_2]_z
    F^{\mathrm{mes}}_{\pi NN} (\vec{p}_1^2,\vec{p}_2^2)
    (\vec{p}_1 \cdot \vecg{\sigma}_1) (\vec{p}_2 \cdot \vecg{\sigma}_2)
    (\vec{p}_1 - \vec{p}_2), \\
    \label{eq:J2nrho}  %
    \vec{j}^{[2]}_{\rho NN}(\vec{Q},\vec{p}_1, \vec{p}_2) = {}&
    -  f_1^{\mathrm{IV}}(Q^2) \big\{ [i \vecg{\tau}_1 \times \vecg{\tau}_2]_z
    F^{\mathrm{con}}_{\rho NN} (\vec{p}_2^2)
    [(\vecg{\sigma}_2 \times \vec{p}_2) \times \vecg{\sigma}_1 ]
    + (1 \leftrightarrow 2) \big \} \nonumber \\
    {}& + f_1^{\mathrm{IV}}(Q^2) [i \vecg{\tau}_1 \times \vecg{\tau}_2]_z
    F^{\mathrm{mes}}_{\rho NN} (\vec{p}_1^2,\vec{p}_2^2)
    [(\vec{p}_1 \times \vecg{\sigma}_1) \cdot (\vec{p}_2 \times \vecg{\sigma}_2)]
    (\vec{p}_1 - \vec{p}_2) \nonumber \\
    {}& + f_1^{\mathrm{IV}}(Q^2) [i \vecg{\tau}_1 \times \vecg{\tau}_2]_z
    F^{\mathrm{mes1}}_{\rho NN} (\vec{p}_1^2,\vec{p}_2^2)
    (\vec{p}_1 - \vec{p}_2) \nonumber \\
    {}& - f_1^{\mathrm{IV}}(Q^2) [i \vecg{\tau}_1 \times \vecg{\tau}_2]_z
    F^{\mathrm{mes}}_{\rho NN} (\vec{p}_1^2,\vec{p}_2^2) \vec{Q} \times
    [(\vec{p}_1 \times \vecg{\sigma}_1) \times (\vec{p}_2 \times\vecg{\sigma}_2)],
    \\ %
    \vec{j}^{[2]}_{\rho\pi NN}(\vec{Q},\vec{p}_1, \vec{p}_2) = {}&
    -  f_1^{\mathrm{IS}}(Q^2) \big\{ ( \vecg{\tau}_1 \cdot \vecg{\tau}_2)
    F^{\mathrm{dis}}_{\rho\pi NN} (\vec{p}_1^2,\vec{p}_2^2)
    (\vec{p}_2 \cdot \vecg{\sigma}_2) [i \vec{p}_1 \times \vec{p}_2]
    + (1 \leftrightarrow 2) \big \}, \\
    \vec{j}^{[2]}_{\omega\pi NN}(\vec{Q},\vec{p}_1, \vec{p}_2) = {}&
    -  f_1^{\mathrm{IV}}(Q^2) \big\{ {\tau}_{2\,z}
    F^{\mathrm{dis}}_{\omega\pi NN} (\vec{p}_1^2,\vec{p}_2^2)
    (\vec{p}_2 \cdot \vecg{\sigma}_2) [i \vec{p}_1 \times \vec{p}_2]
    + (1 \leftrightarrow 2) \big \} ,
  \end{align}
\end{subequations}
\begin{subequations} \label{eq:J2nd}
  \begin{align} 
    \vec{j}^{[2]}_{\pi \Delta N}(\vec{Q},\vec{p}_1, \vec{p}_2) = {}&
    -  f_1^{\mathrm{IV}}(Q^2) \big\{ [i \vecg{\tau}_1 \times \vec{T}_{2}]_z
    F^{\mathrm{con}}_{\pi \Delta N} (\vec{p}_2^2)
    (\vec{p}_2 \cdot \vec{S}_2) \vecg{\sigma}_1 
    + (1 \leftrightarrow 2) \big \} \nonumber \\
    {} & -  f_1^{\mathrm{IV}}(Q^2) \big\{ [i \vec{T}_1 \times \vecg{\tau}_{2}]_z
    F^{\mathrm{con}}_{\pi \Delta N} (\vec{p}_2^2)
    (\vec{p}_2 \cdot \vecg{\sigma}_{2})
    \vec{S}_1 + (1 \leftrightarrow 2) \big \} \nonumber \\
    {}& + f_1^{\mathrm{IV}}(Q^2) \big\{ [i \vecg{\tau}_1 \times \vec{T}_2]_z
    F^{\mathrm{mes}}_{\pi \Delta N} (\vec{p}_1^2,\vec{p}_2^2)
    (\vec{p}_1 \cdot \vecg{\sigma}_1) (\vec{p}_2 \cdot \vec{S}_2)
    (\vec{p}_1 - \vec{p}_2)  + (1 \leftrightarrow 2) \big \} ,  \\
    \vec{j}^{[2]}_{\rho \Delta N}(\vec{Q},\vec{p}_1, \vec{p}_2) = {}&
    -  f_1^{\mathrm{IV}}(Q^2) \big\{ [i \vecg{\tau}_1 \times \vec{T}_2]_z
    F^{\mathrm{con}}_{\rho \Delta N} (\vec{p}_2^2)
    [(\vec{S}_2 \times \vec{p}_2) \times \vecg{\sigma}_1 ]
    + (1 \leftrightarrow 2) \big \} \nonumber \\
    {}& - f_1^{\mathrm{IV}}(Q^2) \big\{ [i \vec{T}_1 \times \vecg{\tau}_2]_z
    F^{\mathrm{con}}_{\rho \Delta N} (\vec{p}_2^2)
    [(\vecg{\sigma}_2 \times \vec{p}_2) \times \vec{S}_1 ]
    + (1 \leftrightarrow 2) \big \} \nonumber \\
    {}& + f_1^{\mathrm{IV}}(Q^2) \big\{ [i \vecg{\tau}_1 \times \vec{T}_2]_z
    F^{\mathrm{mes}}_{\rho \Delta N} (\vec{p}_1^2,\vec{p}_2^2)
    [(\vec{p}_1 \times \vecg{\sigma}_1) \cdot (\vec{p}_2 \times \vec{S}_2)]
    (\vec{p}_1 - \vec{p}_2) + (1 \leftrightarrow 2) \big \}  \nonumber \\
    {}& - f_1^{\mathrm{IV}}(Q^2) \big\{ [i \vecg{\tau}_1 \times \vec{T}_2]_z
    F^{\mathrm{mes}}_{\rho \Delta N} (\vec{p}_1^2,\vec{p}_2^2) \vec{Q} \times
    [(\vec{p}_1 \times \vecg{\sigma}_1) \times (\vec{p}_2 \times\vec{S}_2)]
    + (1 \leftrightarrow 2) \big \},  \\ 
    \vec{j}^{[2]}_{\rho\pi \Delta N}(\vec{Q},\vec{p}_1, \vec{p}_2) = {}&
    -  f_1^{\mathrm{IS}}(Q^2) \big\{ ( \vecg{\tau}_1 \cdot \vec{T}_2)
    F^{\mathrm{dis}}_{\rho\pi \Delta N} (\vec{p}_1^2,\vec{p}_2^2)
    (\vec{p}_2 \cdot \vec{S}_2) [i \vec{p}_1 \times \vec{p}_2]
    + (1 \leftrightarrow 2) \big \}, \\ 
    \vec{j}^{[2]}_{\omega\pi \Delta N}(\vec{Q},\vec{p}_1, \vec{p}_2) = {}&
    -  f_1^{\mathrm{IV}}(Q^2) \big\{ {T}_{2\,z}
    F^{\mathrm{dis}}_{\omega\pi \Delta N} (\vec{p}_1^2,\vec{p}_2^2)
    (\vec{p}_2 \cdot \vec{S}_2) [i \vec{p}_1 \times \vec{p}_2]
    + (1 \leftrightarrow 2) \big \} ,
  \end{align}
\end{subequations}
    
\begin{gather} 
  \label{eq:J2d}
  \begin{align} 
    \vec{j}^{[2]}_{\pi \Delta \Delta}(\vec{Q},\vec{p}_1, \vec{p}_2) = {}&
    -  f_1^{\mathrm{IV}}(Q^2) \big \{ 
    [i \vecg{\tau}_1 \times \vecg{\tau}_{\Delta \,2}]_z
    F^{\mathrm{con, \; d}}_{\pi \Delta \Delta} (\vec{p}_2^2)
    (\vec{p}_2 \cdot \vecg{\sigma}_{\Delta \,2})
    \vecg{\sigma}_1 + (1 \leftrightarrow 2) \big \} \nonumber \\
    {} & -  f_1^{\mathrm{IV}}(Q^2) \big\{ 
    [i \vecg{\tau}_{\Delta \,1} \times \vecg{\tau}_2]_z
    F^{\mathrm{con, \; d}}_{\pi \Delta \Delta} (\vec{p}_2^2)
    (\vec{p}_2 \cdot \vecg{\sigma}_2)
    \vecg{\sigma}_{\Delta \,1} + (1 \leftrightarrow 2) \big \} \nonumber \\
    {}& + f_1^{\mathrm{IV}}(Q^2)  \big\{ 
    [i \vecg{\tau}_1 \times \vecg{\tau}_{\Delta \,2}]_z
    F^{\mathrm{mes, \; d}}_{\pi \Delta \Delta} (\vec{p}_1^2,\vec{p}_2^2)
    (\vec{p}_1 \cdot \vecg{\sigma}_1) (\vec{p}_2 \cdot \vecg{\sigma}_{\Delta \,2})
    (\vec{p}_1 - \vec{p}_2) + (1 \leftrightarrow 2) \big \} \nonumber \\
    {}& -  f_1^{\mathrm{IV}}(Q^2) \big\{ 
    [i \vec{T}_1^{\dagger} \times \vec{T}_{2}]_z
    F^{\mathrm{con,\;e}}_{\pi \Delta \Delta} (\vec{p}_2^2)
    (\vec{p}_2 \cdot \vec{S}_2) \vec{S}_1 ^{\dagger}
    + (1 \leftrightarrow 2) \big \} \nonumber \\
    {} & - f_1^{\mathrm{IV}}(Q^2) \big\{ [i \vec{T}_1 \times \vec{T}_2^{\dagger}]_z
    F^{\mathrm{con,\;e}}_{\pi \Delta \Delta} (\vec{p}_2^2)
    (\vec{p}_2 \cdot \vec{S}_{2}^{\dagger})
    \vec{S}_1 + (1 \leftrightarrow 2) \big \} \nonumber \\
    {}& + f_1^{\mathrm{IV}}(Q^2) \big\{ [i \vec{T}_1^{\dagger} \times \vec{T}_2]_z
    F^{\mathrm{mes,\;e}}_{\pi \Delta \Delta} (\vec{p}_1^2,\vec{p}_2^2)
    (\vec{p}_1 \cdot \vec{S}_1^{\dagger}) (\vec{p}_2 \cdot \vec{S}_2)
    (\vec{p}_1 - \vec{p}_2)  + (1 \leftrightarrow 2) \big \}.
  \end{align}
\end{gather} 
\end{widetext}
We note: The contribution to the two-nucleon $\rho$-exchange current, 
proportional to $F^{\mathrm{mes1}}_{\rho NN} (\vec{p}_1^2,\vec{p}_2^2)$
\Eq~\eqref{eq:J2nrho} is not contained in the standard collection of 
exchange currents of 
\Refs~\cite{yuan:02a,strueve:87a,henning:92a,oelsner:phd}, used by us 
till now in the context of other potentials; it is necessitated in this paper
by the full form of the $\rho$ exchange implemented in the 
CD-Bonn potential. Other contributions arising from the full $\rho$ exchange
\cite{adam:89a} are of higher order compared to 
$F^{\mathrm{mes1}}_{\rho NN} (\vec{p}_1^2,\vec{p}_2^2)$
and therefore  are neglected in our standard calculations;
their effect is discussed in \Sect~\ref{sec:Smec}.

The $F$-functions used in the above expressions are potential-dependent.
For meson-exchange potentials they are built from  
 meson-baryon coupling constants, hadronic form factors and meson propagators.
For contact currents the $F$-functions have the following forms:
\begin{subequations} \label{eq:Fsc}
  \begin{align} 
    F^{\mathrm{con}}_{\pi NN} (\vec{p}^2) = {} & 
    \frac{1}{8\pi^2 m_N^2} \frac{g_{\pi}^2}{4\pi}
    \frac{\mathcal{F}_{\pi NN}^2(\vec{p}^2)}{m_{\pi}^2 + \vec{p}^2}, \\
    F^{\mathrm{con}}_{\rho NN} (\vec{p}^2) = {} & 
    \frac{1}{8\pi^2 m_N^2} \frac{g_{\rho}^2 (1+f_{\rho}/g_{\rho})^2}{4\pi}
    \frac{\mathcal{F}_{\rho NN}^2(\vec{p}^2)}{m_{\rho}^2 + \vec{p}^2}, \\
    F^{\mathrm{con}}_{\pi \Delta N} (\vec{p}^2) = {} & 
    \frac{1}{8\pi^2 m_N^2} \frac{g_{\pi}^2}{4\pi} 
    \frac{f_{\pi N\Delta}}{f_{\pi NN}}
    \frac{\mathcal{F}_{\pi NN}(\vec{p}^2) \mathcal{F}_{\pi \Delta N}(\vec{p}^2)}
    {m_{\pi}^2 + \vec{p}^2}, \\
    F^{\mathrm{con}}_{\rho \Delta N} (\vec{p}^2) = {} & 
    \frac{1}{8\pi^2 m_N^2} \frac{g_{\rho}^2 (1+f_{\rho}/g_{\rho})^2}{4\pi}
    \frac{f_{\rho N\Delta}}{f_{\rho NN}}
    \nonumber \\ & \times
    \frac{\mathcal{F}_{\rho NN}(\vec{p}^2) \mathcal{F}_{\rho \Delta N}(\vec{p}^2)}
    {m_{\rho}^2 + \vec{p}^2}, \\
    F^{\mathrm{con,\;d}}_{\pi \Delta \Delta} (\vec{p}^2) = {} & 
    \frac{1}{8\pi^2 m_N^2} \frac{g_{\pi}^2}{4\pi} 
    \frac{f_{\pi \Delta \Delta}}{f_{\pi NN}} \frac{\mathcal{F}_{\pi NN}(\vec{p}^2) 
      \mathcal{F}_{\pi \Delta \Delta}(\vec{p}^2)}  {m_{\pi}^2 + \vec{p}^2}, \\
    F^{\mathrm{con,\;e}}_{\pi \Delta \Delta} (\vec{p}^2) = {} & 
    \frac{1}{8\pi^2 m_N^2} \frac{g_{\pi}^2}{4\pi} 
    \frac{f_{\pi N \Delta}^2}{f_{\pi NN}^2} 
    \frac{\mathcal{F}_{\pi \Delta N}^2(\vec{p}^2)} {m_{\pi}^2 + \vec{p}^2}.
  \end{align}
\end{subequations}
For meson in flight currents the corresponding expressions are
\begin{subequations} \label{eq:Fsm}
  \begin{align} 
    F^{\mathrm{mes}}_{\alpha B'B} (\vec{p}_1^2,\vec{p}_2^2) = & 
    - \frac{1}{\vec{p}_1^2 - \vec{p}_2^2} 
    \big[ F^{\mathrm{con}}_{\alpha B'B} (\vec{p}_1^2) - 
    F^{\mathrm{con}}_{\alpha B'B} (\vec{p}_2^2) \big], \\
    F^{\mathrm{mes1}}_{\rho NN} (\vec{p}_1^2,\vec{p}_2^2) = {} & 
    \frac{4m_N^2}{(1+f_{\rho}/g_{\rho})^2}
    F^{\mathrm{mes}}_{\rho NN} (\vec{p}_1^2,\vec{p}_2^2), \\
    F^{\mathrm{mes,\;d(e)}}_{\pi \Delta \Delta} (\vec{p}_1^2,\vec{p}_2^2) = {} & 
    - \frac{1}{\vec{p}_1^2 - \vec{p}_2^2} 
    \nonumber \\ & \times 
    \big[ F^{\mathrm{con,\;d(e)}}_{\pi \Delta \Delta} (\vec{p}_1^2) 
    - F^{\mathrm{con,\;d(e)}}_{\pi \Delta \Delta} (\vec{p}_2^2) \big].
  \end{align}
\end{subequations}
Finally, the functions for nondiagonal meson-exchange currents 
 (also called dispersion currents) are defined to be
\begin{subequations} \label{eq:Fsd}
  \begin{align} 
    F^{\mathrm{dis}}_{\alpha\beta N N} (\vec{p}_1^2,\vec{p}_2^2) = {} & 
    \frac{1}{4\pi^2 m_N^2} \frac{g_{\alpha} g_{\beta}}{4\pi}
    \frac{m_N}{m_{\alpha}} g_{\alpha \beta \gamma}
    \nonumber \\ & \times 
    \frac{\mathcal{F}_{\alpha NN}(\vec{p}_1^2)} {m_{\alpha}^2 + \vec{p}_1^2}
    \frac{\mathcal{F}_{\beta NN}(\vec{p}_2^2)} {m_{\beta}^2 + \vec{p}_2^2}, \\
    F^{\mathrm{dis}}_{\alpha\beta \Delta N} (\vec{p}_1^2,\vec{p}_2^2) = {} & 
    \frac{1}{4\pi^2 m_N^2} \frac{g_{\alpha} g_{\beta}}{4\pi}
    \frac{f_{\pi N\Delta}}{f_{\pi NN}} \frac{m_N}{m_{\alpha}} 
    g_{\alpha \beta \gamma}
    \nonumber \\ & \times   
    \frac{\mathcal{F}_{\alpha NN}(\vec{p}_1^2)} {m_{\alpha}^2 + \vec{p}_1^2}
    \frac{\mathcal{F}_{\beta \Delta N}(\vec{p}_2^2)} {m_{\beta}^2 + \vec{p}_2^2}.
  \end{align}
\end{subequations}
The meson-nucleon coupling constants $g_{\alpha}$ and $f_{\rho}$ are listed in 
Table I of \Ref~\cite{machleidt:01a}, whereas other hadronic parameters,
i.e., coupling constants $f_{\alpha B'B}$, meson masses $m_{\alpha}$
and hadronic form factors $\mathcal{F}_{\alpha B'B}(\vec{p}^2)$ are those of
\Ref~\cite{deltuva:03c}. The e.m. meson-photon coupling constants have the
values $g_{\rho \pi \gamma} = 0.56$ and $g_{\omega \pi \gamma} = 0.68$
according to \Ref~\cite{carlson:98a}.

\subsection{Operator corrections of lowest relativistic order}

Sample operator corrections of relativistic order for the charge density are given.
They are of one-baryon and of two-baryon nature:
\begin{subequations} \label{eq:Crel}
  \begin{align} 
    \label{eq:c1nrc}
    \rho^{[1] \; \mathrm{rc}}_{NN } (\vec{Q},\vec{k}',\vec{k}) = {}&
    - \frac{e(Q^2) + 2\kappa(Q^2)}{8 m_N^2}
  \nonumber \\ & \times 
    \big \{ \vec{Q}^2 + [i\vecg{\sigma} \times (\vec{k}'+ \vec{k})] 
    \cdot \vec{Q} \big \}, \\
    \rho^{[1] \; \mathrm{rc}}_{\Delta N} (\vec{Q},\vec{k}',\vec{k}) = {}&
    -\frac{1}{4 m_N m_{\Delta} }   g^{\mathrm{M1}}_{\Delta N}(Q^2) 
  \nonumber \\ & \times 
    [i\vec{S} \times (\vec{k}'+ \vec{k})] \cdot \vec{Q} \; T_z, 
    \\  \label{eq:Crel2N}
    \rho^{[2]\; \mathrm{rc}}_{\pi NN}(\vec{Q},\vec{p}_1, \vec{p}_2) = {}& 
    \frac{1}{2m_N} [f_1^{\mathrm{IS}}(Q^2) \vecg{\tau}_1 \cdot \vecg{\tau}_2 
      + f_1^{\mathrm{IV}}(Q^2) \tau_{2\,z}]
  \nonumber \\ & \times 
    F^{\mathrm{con}}_{\pi NN} (\vec{p}_2^2) (\vecg{\sigma}_1 \cdot \vec{Q}) \,
    (\vecg{\sigma}_2 \cdot \vec{p}_2) 
    \nonumber \\ & 
    + (1 \leftrightarrow 2).
  \end{align}
\end{subequations}
The contributions \eqref{eq:Crel} are the Darwin-Foldy and spin-orbit
corrections of the one-nucleon charge density, the one-baryon  correction
due to nucleon-$\Delta$ transition and the two-nucleon correction due to
$\pi$ exchange, respectively; the two-nucleon contribution \eqref{eq:Crel2N}
is local and therefore often exclusively used; there are however
other nonlocal two-nucleon contributions of the same order.
The contributions \eqref{eq:Crel} are used in this paper in 
\Sect~\ref{sec:rel} for the Siegert form of the current.
 Since they are relativistic corrections, they violate current conservation
in the considered order. However, the calculated trinucleon elastic charge 
form factors need all three contributions in order to become almost quantitatively 
consistent with the experimental data~\cite{henning:92a}.

\section{\label{sec:inteq} Integral equation for current matrix element}

This appendix calculates the current matrix elements of two- and
three-body photo disintegration of the trinucleon bound state, i.e., 
$\langle\psi^{(-)}_{\alpha} (\vec{q}_f) \nu_{\alpha_f}| 
j^{\mu} (\vec{k}_{\gamma}, \Kpl)
\epsilon_{\mu} (\vec{k}_{\gamma} \lambda) | B \rangle $ and
$\langle\psi^{(-)}_{0} (\vec{p}_f \vec{q}_f) \nu_{0_f}| 
j^{\mu} (\vec{k}_{\gamma}, \Kpl)
\epsilon_{\mu} (\vec{k}_{\gamma} \lambda) | B \rangle $.

The antisymmetrized fully correlated three-nucleon
scattering states of internal motion in nucleon-deuteron channels, i.e.,
$\langle\psi^{(-)}_{\alpha} (\vec{q}_f) \nu_{\alpha_f}|$, and
in three-body breakup channels, i.e.,
$\langle\psi^{(-)}_{0} (\vec{p}_f \vec{q}_f) \nu_{0_f}|$, are
 not calculated explicitly; they are calculated only implicitly when forming 
current matrix elements.  We introduce the 
state $|X(Z) \rangle$, defined according to
\begin{subequations} \label{eq:X}
  \begin{align} \label{eq:Xa}
    |X(Z) \rangle = {} &  \big( 1+P \big) 
    j^{\mu} (\vec{k}_{\gamma}, \Kpl) 
    \epsilon_{\mu} (\vec{k}_{\gamma} \lambda) | B \rangle
    \nonumber \\ & 
    + P T(Z) G_0(Z) |X(Z) \rangle, \\  \label{eq:Xb}
    |X (Z)\rangle = {} & \sum_{n=0}^{\infty} [P T(Z) G_0(Z)]^n
    \nonumber \\ & \times 
    \big( 1+P \big) j^{\mu} (\vec{k}_{\gamma}, \Kpl) 
    \epsilon_{\mu} (\vec{k}_{\gamma} \lambda) | B \rangle,
  \end{align} 
\end{subequations}
as intermediate quantity
with $Z=E_i+i0$ being the three-particle available energy and $T(Z)$ being
the two-baryon transition matrix. 
Equation~\eqref{eq:Xa} is an integral equation for  $|X (Z)\rangle$, analogous
to that for the multichannel transition matrix $U(Z)$ of \Ref~\cite{deltuva:03a}:
Both equations have the same kernel, only their driving terms are different.
We therefore solve \Eq~\eqref{eq:Xa} according to the technique of 
\Ref~\cite{deltuva:03a}, summing the Neumann series~\eqref{eq:Xb} for
$|X (Z)\rangle$ by the Pad\'e method. Once $|X (Z)\rangle$ is calculated, 
the current matrix elements required for the description of two- and
three-body photo disintegration of the trinucleon bound state are obtained
according to 
\begin{subequations} \label{eq:X2J}
  \begin{align} \label{eq:X2Ja}
    \langle  \psi^{(-)}_{\alpha} &(\vec{q}_f) \nu_{\alpha_f}| 
    j^{\mu} (\vec{k}_{\gamma}, \Kpl)
    \epsilon_{\mu} (\vec{k}_{\gamma} \lambda) | B \rangle  
\nonumber \\  = {} & 
    { \frac{1}{\sqrt{3}} }
    \langle \phi_{\alpha} (\vec{q}_f) \nu_{\alpha_f}| X (Z)\rangle, 
    \\  \nonumber
    \langle  \psi^{(-)}_{0} & (\vec{p}_f \vec{q}_f) \nu_{0_f}| 
    j^{\mu} (\vec{k}_{\gamma}, \Kpl)
    \epsilon_{\mu} (\vec{k}_{\gamma} \lambda) | B \rangle \\ = {} & 
    { \frac{1}{\sqrt{3}} }
    \langle \phi_{0} (\vec{p}_f \vec{q}_f) \nu_{0_f}| 
    \big( 1+P \big) 
    \big[ j^{\mu} (\vec{k}_{\gamma}, \Kpl ) 
    \epsilon_{\mu} (\vec{k}_{\gamma} \lambda) | B \rangle  
    \nonumber \\ & 
    + T(Z) G_0(Z) |X (Z)\rangle \big].  \label{eq:X2Jb}
  \end{align} 
\end{subequations}
The current matrix element required for the description of radiative 
nucleon-deuteron capture is related to that of two-body photo disintegration
by time reversal as described in \Ref~\cite{yuan:02a}.
When calculating total two- and three-body photo disintegration cross section
$\sigma$, the integration over all final states can be performed implicitly, i.e., 
\begin{subequations} \label{eq:sigtot}
  \begin{align} \label{eq:sigtota}
    \sigma = &{} \frac{(2\pi\hbar)^2 }{ \hbar c^2 k_{\gamma}^0} \,
    \frac14 \sum_{\mathcal{M}_B \lambda}
    \langle B |[j^{\mu} (\vec{k}_{\gamma}, \Kpl) 
    \epsilon_{\mu} (\vec{k}_{\gamma} \lambda)]^{\dagger}
\nonumber \\ & \times 
    \delta (E_i - H_0 - H_I)    j^{\mu} (\vec{k}_{\gamma}, \Kpl) 
    \epsilon_{\mu} (\vec{k}_{\gamma} \lambda) | B \rangle, 
\end{align}
  \begin{align}    \label{eq:sigtotb}
    \sigma = &{}  -\frac{(2\pi\hbar)^2 }{4 \pi \hbar c^2 k_{\gamma}^0}
    \sum_{\mathcal{M}_B \lambda} \mathrm{Im}  \Big\{
    \langle B |[j^{\mu} (\vec{k}_{\gamma}, \Kpl) 
    \epsilon_{\mu} (\vec{k}_{\gamma} \lambda)]^{\dagger}
\nonumber \\ & \times 
    G(\Ei)    j^{\mu} (\vec{k}_{\gamma}, \Kpl) 
    \epsilon_{\mu} (\vec{k}_{\gamma} \lambda) | B \rangle \Big\}.
  \end{align}
The auxiliary state  $G(\Ei)  j^{\mu} (\vec{k}_{\gamma}, \Kpl) 
    \epsilon_{\mu} (\vec{k}_{\gamma} \lambda) | B \rangle $
of \Eq~\eqref{eq:sigtotb} is related to  $|X(\Ei) \rangle$ according to 
\begin{gather} \label{eq:sigtotc}
  \begin{align} 
    G(\Ei) & j^{\mu} (\vec{k}_{\gamma}, \Kpl) 
    \epsilon_{\mu} (\vec{k}_{\gamma} \lambda) | B \rangle
\nonumber \\ = {} &  
    \frac13 (1+P) G_0(\Ei) \big[ j^{\mu} (\vec{k}_{\gamma}, \Kpl) 
	   \epsilon_{\mu} (\vec{k}_{\gamma} \lambda) | B \rangle  \nonumber \\ 
	   & +  T(\Ei) G_0(\Ei) |X (\Ei)\rangle \big].
  \end{align} 
\end{gather}
The total lab cross section is then obtained in the form
\begin{gather} \label{eq:sigtotd}
  \begin{align} 
  \sigma = &{}  - \frac{(2\pi\hbar)^2 }{12 \pi \hbar c^2 k_{\gamma}^0} \, 
  \sum_{\mathcal{M}_B \lambda}   \mathrm{Im}
  \Big\{ \langle B |[j^{\mu} (\vec{k}_{\gamma}, \Kpl ) 
    \epsilon_{\mu} (\vec{k}_{\gamma} \lambda)]^{\dagger}
\nonumber \\ & \times 
  (1+P) G_0(\Ei) \big[ j^{\mu} (\vec{k}_{\gamma}, \Kpl) 
	   \epsilon_{\mu} (\vec{k}_{\gamma} \lambda) | B \rangle
\nonumber \\ & 
	   +  T(\Ei) G_0(\Ei) |X (\Ei)\rangle \big] \Big\}.
  \end{align} 
\end{gather}
\end{subequations}
We note that \Eqs~\eqref{eq:sigtot} performs the integration over all final
states implicitly using the nonrelativistic Hamiltonian in contrast to the 
strategy of \Eq~\eqref{eq:dLips}.

\end{appendix}

\end{document}